\documentclass[aps,prl,showpacs,twocolumn,preprintnumbers,groupedaddress]{revtex4-1}
\usepackage{charter,graphicx,amssymb,amsmath,amsfonts,mathtools}
\usepackage[colorlinks=true,citecolor=blue]{hyperref}
\usepackage{mathpazo,times}
\usepackage{epstopdf}

\let\eref=\eqref
\def\gl{\mathrel{\mathpalette\overl@ss>}}
\def\lg{\mathrel{\mathpalette\overgr@at<}}
\def\d{\mathrm{d}}
\def\Natural{\mathbb{N}}
\def\Integer{\mathbb{Z}}
\def\Real{\mathbb{R}}
\def\Complex{\mathbb{C}}
\def\Re{\mathop{\rm Re}\nolimits}
\def\Im{\mathop{\rm Im}\nolimits}
\def\arg{\mathop{\rm arg}\nolimits}
\def\pvint{\mathop{\int\kern-0.94em-\kern0.2em}\limits}
\let\^=\hat
\let\==\bar
\let\@=\mathbf
\let\letrue=\le
\let\getrue=\ge
\let\le=\leqslant
\let\ge=\geqslant
\def\d{\mathrm{d}}
\def\e{\mathrm{e}}
\def\o{\mathrm{o}}
\def\Wr{\mathrm{Wr}}
\def\tr{\mathrm{tr}}
\def\sech{\mathrm{sech}}

\def\O#1{^{(#1)}}
\def\~#1{\tilde{#1}}
\def\[{\begin{equation}}
\def\]{\end{equation}}
\def\bse{\begin{subequations}}
\def\ese{\end{subequations}}

\advance\abovedisplayskip -10pt
\advance\abovedisplayshortskip -10pt
\advance\belowdisplayskip 2pt
\advance\belowdisplayshortskip 2pt
\advance\parskip 0pt
\allowdisplaybreaks

\def\blue#1{\textcolor{black}{#1}}
\def\beginblue{\begingroup\color{black}}
\def\endblue{\endgroup}

\begin{document}
\title{
Recurrence due to periodic multi-soliton fission in the defocusing nonlinear Schr\"odinger equation
%How universal is the recurrence due to periodic multi-soliton fission?
}
\author{Guo Deng$^1$}
\author{Sitai Li$^2$}
\author{Gino Biondini$^{1,2}$}
\author{Stefano Trillo$^3$}
\affiliation{
  $^1$Department of Physics, State University of New York at Buffalo, Buffalo, New York 14260, USA\\
  $^2$Department of Mathematics, State University of New York at Buffalo, Buffalo, New York 14260, USA\\
  $^3$Department of Engineering, University of Ferrara, Via Saragat 1, 44122 Ferrara, Italy
}
\date{\today}

%%%%%%%%%%%%%%%%%%%%%%%%%%%%%%%%%%%%%%%%%%%%%%%%%%%%%%%%%%%%%%%%%%%%%%%%%%%%%%%%%%%%%%%%%%%%%
\begin{abstract}
We address the degree of universality of the Fermi-Pasta-Ulam recurrence induced by multisoliton fission from a harmonic excitation by analysing the case of the semiclassical defocusing nonlinear Schr\"odinger equation,
{which models nonlinear wave propagation in a variety of physical settings. Using}
a suitable Wentzel-Kramers-Brillouin approach to the solution of the associated scattering problem {we} accurately predict, in full analytical way, the number and the features (amplitude and velocity) of soliton-like excitations emerging post-breaking, as a function of the dispersion smallness parameter.
This also permits to predict and analyse the near-recurrences, thereby inferring the universal character of the mechanism originally discovered for the Korteweg-deVries equation.
%[Phys. Rev. Lett. 15, 240 (1965)].
We show, however, that important differences {exist between} the two models, {arising} from the different scaling rules obeyed by the soliton velocities.\\
%\end{abstract}

%\begin{abstract}
%We investigate the broad nature of small-dispersion phenomena, including hydrodynamic instability and wave breaking, soliton generation and Fermi-Pasta-Ulam recurrence, by analyzing these problems in the context of the semiclassical defocusing nonlinear Schr\"odinger (NLS) equation and by comparing the corresponding results with those for the Korteweg-deVries equation. By employing a Wentzell-Kramers-Brillouin approximation in the associated scattering problem for the NLS equation, we analytically characterize the post-breaking dynamics by predicting the number of fissioning solitons as well as their features (amplitude and velocity) as a function of the dispersion smallness parameter. In turn, this allows us to describe the recurrence phenomenon, calculate analytically the recurrence times and characterize the degree of recurrence. Comparison with the KdV equation reveals important quantitative differences in the recurrence phenomena in the two models, due to the different scaling rules obeyed by the
%soliton velocities in the two models.
\end{abstract}
%\\[0.4ex]
%PACS:
%02.30.Ik, % Integrable systems
%05.45.-a, % Nonlinear dynamics and chaos
%05.45.Yv, % Solitons
%42.65.Sf, % Dynamics of nonlinear optical systems; optical instabilities, optical chaos and complexity, and optical spatio-temporal dynamics
%47.20.-k  % Flow instabilities

\pacs{02.30.Ik, 05.45.Yv, 42.65.Sf, 47.20.-k}
%Keywords:
%Nonlinear Schr\"odinger systems, semiclassical limits, hydrodynamic instabilities, solitons, recurrence.
%\keywords{Nonlinear Schr\"odinger systems, semiclassical limits, hydrodynamic instabilities, solitons, recurrence}%Use showkeys class option if keyword display desired

\maketitle

%%%%%%%%%%%%%%%%%%%%%%%%%%%%%%%%%%%%%%%%%%%%%%%%%%%%%%%%%%%%%%%%%%%%%%%%%%%%%%%%%%%%%%%%%%%%%
\section{\noexpand\blue{I.~ Introduction}}

The {discovery} by Fermi-Pasta-Ulam (FPU) that a low frequency excitation of nonlinear oscillator chains give rise to recurrence instead of equipartition,
a phenomenon now known as FPU recurrence \cite{fpu1955,Tuck1972}, turned into one of most {consequential findings} of nonlinear physics \cite{Ford1992,Berman2005,Dauxois2005,Gallavotti2008,FPU2009divulgative},
{leading to multiple research avenues which are}
still actively {investigated} today \cite{Benettin2013,Onorato2015,Weiner2016,Guasoni2017}.
The first resolution of the {apparent} paradox was given ten years later by Zabusky and Kruskal (ZK) \cite{ZK} who showed that in the continuum limit that gives rise to the weakly dispersive Korteweg-de Vries (KdV) equation \cite{KdV1895}, the recurrence can be understood in terms of solitons, that fission from points of breaking (shocks) occurring in the periodic input mode (a regime which is substantially confirmed, e.g. for the quadratic, or so-called $\alpha-$type, FPU chain \cite{Lorenzoni06}).
Nearly synchronous arrival of the solitons after interaction leads to the recursion.
An explicit estimate for the KdV recurrence time was then given by Toda~\cite{toda1,toda2}.
\blue{Experimentally, the recurrence from periodic input was reported in different systems ranging from electrical lattice networks \cite{Hirota70} to continuous wave systems such as ion acoustic plasma waves
\cite{Ikezi73}, and gravity waves in shallow water \cite{prl2016deng}, while it could be potentially observed also for electron beams that exhibit KdV type of breaking \cite{Mo13}.
}
%Later on, Toda explained the linear variation of the velocity of the largest soliton of the same setting, and gave an explicit estimate for the recurrence time~\cite{toda1,toda2}.
%Experimental observation investigation  have been carried out in systems described by the KdV \cite{Ikezi73}, until recently when recurrence was finally observed for shallow water waves \cite{prl2016deng}.

\blue{
\blue{This} result has led to the \blue{notion} that a time scale exists for which the dynamics of FPU is essentially integrable, that is, it remains close to integrable limits, either the KdV in the continuum approximation or the discrete Toda chain \cite{Benettin2013,Ferguson82}.
Over much longer times the FPU chains eventually thermalise, as it is now clear after decades of research on this topics  \cite{Benettin2013,Onorato2015}.
A crucially important remaining issue, however, is to investigate the metastable state characterised by the recurrences at intermediate {time} scales.
In this context, a natural question that has remained surprisingly unaddressed is {\em the degree of universality of the mechanism discovered by ZK for other integrable models.}%
}
%in addition to KdV.
While it is now accepted that FPU recurrence is not necessarily a prerogative of integrable models \cite{Guasoni2017}, the latter constitutes a wide and extremely important class employed to describe innumerous physical situations.
\blue{Clearly,}
\blue{the importance of integrable systems goes well beyond the specific interest for FPU chains,
\blue{since they often provide} an accurate description of several genuinely continuous physical systems.
%\blue{(in particular, the latter encompass, besides the integrable limit of FPU chains, the physics of several genuinely continuous systems).}
In this context,}
breaking phenomena are a universal feature of the weakly dispersive regime \cite{Ercolani1994} (see also \cite{ElHoefer16} for a recent review) occurring for the nonlinear Schr\"odinger (NLS) equation \cite{Dutton01,Hoefer06,Wan07,Gofra07,Conti09,trillovaliani,fatometrillo,Moro14,Xu17,spinwaves}, naturally arising as the continuum limit of cubic $\beta$-FPU \cite{Berman2005}, the Benjamin-Ono equation \cite{BO,Garnier13} and the Toda lattice \cite{Todashock,Todashock2,Todashock3}, to name only a few.
A special {role in this context is played by} the {\em defocusing} or repulsive NLS equation, since it describes wave breaking phenomena recently observed and analysed in areas as different as nonlinear optics \cite{Wan07,Gofra07,Conti09,trillovaliani,fatometrillo,Moro14,Xu17}, Bose-Einstein condensates \cite{Dutton01,Hoefer06}, and spin waves \cite{spinwaves}.
Furthermore, the periodic input problem for such model was recently demonstrated to be accessible in fiber optics experiments, which show fission of dark ``solitons" from periodic points of breaking \cite{trillovaliani,fatometrillo}.
Yet, natural and fundamental questions concerning how to predict the number and the features of the solitons and whether recurrence should be expected remain to date completely open in such model.

The purpose of this paper is to give an answer to these questions and to demonstrate that soliton generation as a result of hydrodynamic-like instabilities, and the resulting FPU-like recurrences, are in fact a more general feature of nonlinear wave evolution equations with small dispersion.
To this end, we show that such phenomena can be effectively treated analytically using the scattering problem and its finite gap formulation associated with the defocusing NLS equation.
While our results significantly broaden the universality of the ZK mechanism of recurrence, they allow to clarify that the phenomenon presents, for the NLS equation, considerable differences with respect to the case of the KdV equation.

\blue{
We remark that FPU recurrences \blue{have also been} referred to in the literature \blue{in} the focusing regime of \blue{the} NLS equation. These, however, are exact (instead of near) recurrences that occur through a different kind of mechanism involving  modulation instability of a strong background (a forbidden regime for the defocusing NLS equation which is well known to be modulationally stable).
Such \blue{scenarios} have been investigated in fluids and optics only in a regime that involve few dominant Fourier modes, i.e. far from the weakly dispersive limit that we are interested in \cite{lakeyuen,yuenlake,AEK85,TW91,Haelterman,Kimmoun}. Conversely, wavepackets without background in the weakly dispersing limit of the focusing regime are known to exhibit a different type of breaking (i.e., elliptic unbilic catastrophe \cite{DGK09}) compared with the defocusing case. The complicated dynamics in the evolution stage beyond the catastrophe cannot be reduced, in general, to the fission of solitons that simply move apart with
different velocities \cite{MK98,KMP03}. Moreover the global scenario for a periodic input mode in the focusing regime was never addressed so far, and will require a specific analysis that will be proposed elsewhere.
}

\begin{figure*}[t!]
\includegraphics[width=0.995\textwidth]{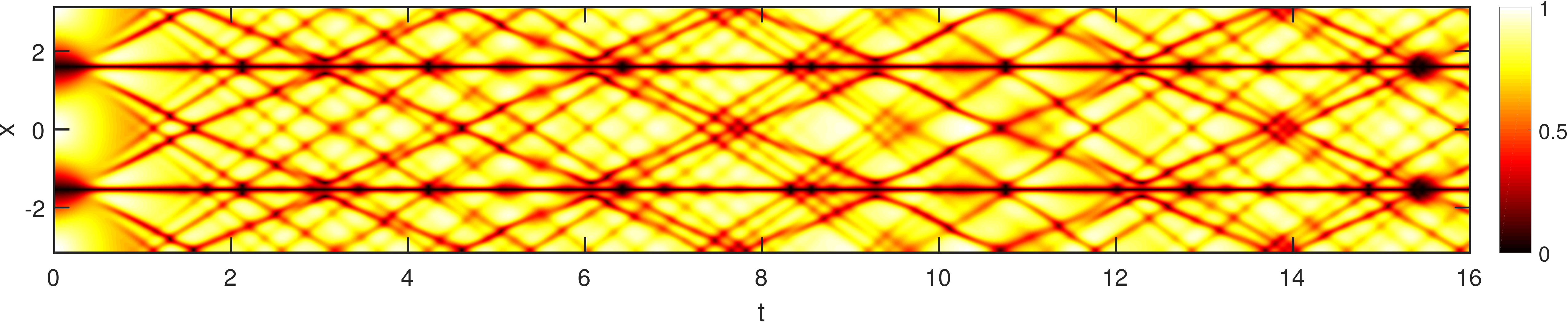}
\caption{Density plot of $|q(x,t)|$ from numerical solution of the initial value problem in Eqs. (\ref{nls}--\ref{ic}) with $\epsilon=0.1$, showing the fission into soliton-like excitations as a result of breaking (hydrodynamic-like instability, see early stage $t <1$) occurring in the null points of the input cosine and the following near-recurrence at $t \simeq15.5$.
%Note that only half of the fundamental spatial domain is plotted since $|q(-x,t)|=|q(x,t)|$.
}
\label{density}\end{figure*}

Motivated by the above discussion, here we consider the following dimensionless {\em periodic} initial value problem for the
\textit{defocusing} NLS equation
\begin{equation}
i\epsilon q_t + \epsilon^2q_{xx} - 2|q|^2q=0\,,
\label{nls}
\end{equation}
with initial conditions
\begin{equation}
q(x,0) = \cos x\,,\qquad -\pi \le x \le \pi\,.
\label{ic}
\end{equation}
The variables $x$ and $t$ are typically space and time, respectively \blue{(e.g., see~\cite{AS1981,InfeldRowlands})},
though in nonlinear fiber optics they have the reversed role of retarded time and propagation distance
\cite{Agrawal,trillovaliani,fatometrillo},
%(up to trivial rescaling) \cite{trillovaliani,fatometrillo}.
and $\epsilon$ quantifies the smallness of dispersion \cite{trillovaliani} (in quantum-mechanical settings $\epsilon$ is also proportional to Planck's constant $\hbar$ \cite{Dutton01}) {compared to the strength of nonlinear effects.}
Without loss of generality \blue{(taking advantage of the scaling invariances of Eq.~\eqref{nls})}, we \blue{have} normalized the period $X_p$ of the cosine initial value to $2\pi$.

The nonlinearity in Eq.~(\ref{nls}) induces conversion towards high-frequency modes, i.e. odd harmonics $\pm m/X_p$, $m=3,5,\ldots$ of the $m=1$ input frequencies, a phenomenon commonly known in nonlinear optics as multiple four-wave mixing (mFWM) \cite{fatometrillo,Thompson91,Trillo94}.
In the regime of interest here, $\epsilon\ll1$, mFWM becomes very efficient and causes strong steepening of the cosine fronts, until breaking (hydrodynamic instability \cite{trillovaliani}) occurs followed by fission into soliton-like excitations, as shown by the numerical simulation in Fig. \ref{density}.
Marked differences with the corresponding phenomena in the KdV equation \cite{ZK,prl2016deng} {are} the degenerate mechanism for breaking that occurs at the null (vacuum) points (cf. \cite{Conti09,Moro14}, where similar mechanism is analysed in detail for dark type input on a constant background) and the fission into pairs with opposite velocities, {which} reflect the bidirectional (NLS) versus unidirectional (KdV) dispersive hydrodynamic nature of such models \cite{ElHoefer16}.
Nevertheless, {even in the NLS equation}, solitons eventually give rise to a near-recurrence as shown in Fig. \ref{density}. Our goal, here, is to {provide a quantitative description of this phenomenon, including}
analytical estimates of the number of fissioning solitons and their velocities as a function of $\epsilon$,
and then use these results to characterise the {recurrent behavior}.

%The experiments in \cite{fatometrillo} showed that a smooth cosine-shaped input input becomes steeper upon propagation,  resulting in the generation of a soliton-like train, in a similar fashion as what happens for the KdV equation.
%No analytical description of the resulting behavior was available until now, however. The scope of the present Letter is to present a complete description of this phenomenon.

%In particular, using a WKB approximation in the scattering problem associated to Eq.~\eqref{nls} with Eq.~\eqref{ic} we  provide a complete asymptotic description of the spectrum, which allows us to completely describe  the number, amplitude and velocities of the emerging solitons for any given value of $\epsilon$. We then use these results to obtain a theoretical estimate of the recurrence time, and we compare the analytical predictions with the results from careful numerical simulations. Finally, we compare the results with those for the KdV equation.

\blue{The outline of this work is the following.
In section~II we characterize the spectrum of the NLS equation~\eqref{nls} in the small dispersion limit
with initial conditions~\eqref{ic} using a suitable WKB expansion.
In section~III we study the properties of the effective solitons arising from the corresponding spectrum.
In section~IV we discuss the recurrence of initial conditions.
Section~V offers a few concluding remarks.
The details of the calculations are given in the Appendix.
}

%%%%%%%%%%%%%%%%%%%%%%%%%%%%%%%%%%%%%%%%%%%%%%%%%%%%%%%%%%%%%%%%%%%%%%%%%%%%%%%%%%%%%%%%%%%%%
\section{\noexpand\blue{II.~ NLS spectrum in the small dispersion limit}}

%%%%%%%%%%%%%%%%%%%%%%%%%%%%%%%%%%%%%%%%%%%%%%%%%%%%%%%%%%%%%%%%%%%%%%%%%%%%%%%%%%%%%%%%%%%%%
\subsection{II.1~ {Scattering problem and monodromy matrix}}

Since the NLS Eq.~(\ref{nls}) is completely integrable, the initial value problem can be solved via the inverse scattering transform (IST).
In particular, the relevant formalism here is the IST with periodic boundary conditions, or finite-gap theory ~\cite{itskotlyarov,MaAblowitz,bbeim,West93}.
According to the periodic IST, the nonlinear excitations embedded in the initial datum are encoded in the spectrum of the scattering problem associated with the NLS equation~(\ref{nls}), i.e. the well-known Zakharov-Shabat (ZS) with cosine potential
%When the IC $q(x,0)$ is given by Eq.~\eref{e:ic}, the scattering problem becomes the semiclassical Zakharov-Shabat (ZS) system
\vspace*{-0.4ex}
\[
\epsilon \phi_x=(-i k\sigma_3+\cos x\,\sigma_1)\,\phi\,,
\label{e:zs}
\]
where $\phi(x,k) = (\phi_1,\phi_2)^T$ is the vector eigenfunction, and
$\sigma_1$ and $\sigma_3$ are the first and third Pauli matrices, respectively.
Since the scattering problem~\eref{e:zs} is self-adjoint, all eigenvalues $k$ are real.
Applying the change of variables $v = \phi_1 + \phi_2$ and $\~v = \phi_1 - \phi_2$,
Eq.~\eref{e:zs} can be reduced to the second-order ordinary differential equation,
\vspace*{-0.4ex}
\[
- \epsilon^2v_{xx}+(\cos^2x - \epsilon\sin x)\,v = \lambda\,v\,,
\label{e:schrodinger}
\]
%where we introduced the shorthand notations $Q(x)=\lambda-\cos^2 x$ and $\lambda=k^2$.
%
%From now on, we will denote Eq.~\eref{e:schrodinger} instead of the Zakharov-Shabat system as the scattering problem,
%and we will use the variable $\lambda$ instead of $k$ {as the spectrum variable}.
which is the time-independent Schr\"odinger equation with an $\epsilon$-dependent potential and eigenvalue $\lambda = k^2\ge0$.
%Since $k\in\Real$, we have $\lambda\ge0$.
Since the potential is periodic, Bloch-Floquet theory can be used to show that
Eq.~\eref{e:schrodinger} admits bounded solutions if and only if
\[
-2\le\tr M\le2\,,
\label{e:trMrange}
\]
where $M(\lambda)$ is the monodromy matrix of the problem, defined as $M=Y^{-1}(-\pi)Y(\pi)$,
and $Y(x)$ is any fundamental matrix solution of Eq.~\eref{e:schrodinger}.
The values of $\lambda$ for which Eq.~\eref{e:trMrange} is satisfied comprise the spectrum of Eq.~\eref{e:schrodinger}.
Note that, to each {nonzero} value of $\lambda$, there correspond two values $k = \pm\sqrt\lambda$ in the original scattering problem~\eref{e:zs}.

\subsection{II.2~ {Wentzel-Kramers-Brillouin (WKB) analysis}}
Since no solutions in closed form are available for Eq.~\eref{e:schrodinger}, we apply the WKB method (e.g., see \cite{Messiah})
to obtain asymptotic expansions for the solutions of Eq.~\eref{e:schrodinger} and therefore for $\tr M$
in the dispersionless limit, similarly to the approach developed in Refs.~\cite{physd2016deng,prl2016deng}.
It is convenient to introduce the shorthand notation
\vspace*{-0.8ex}
\[
Q(x) = \lambda - \cos^2x\,.
\label{e:Qdef}
\]
For $\lambda>1$, no turning points are present, and straightforward calculations
\blue{(see Appendix~A.1 for details)}
yield
\bse
\label{e:trM}
\vspace*{-0.2ex}
\[
\tr M =
2\cos\big(S_0(\lambda)/\epsilon\big),\qquad \lambda>1\,,
\label{e:trM_1}
\]
where $S_0(\lambda) =\int\nolimits_{-\pi}^{\pi} \sqrt{Q(x)}\,dx$.
On the other hand, for $0<\lambda<1$ four turning points are present, and the WKB analysis is considerably more complicated.
\blue{In this case, the main difficulty comes from the fact that one}
\blue{must first construct asymptotic solutions in each of the regions away from the turning points and near each turning points,
and then match the solutions in the transition regions around the turning points,
obtaining so-called connection formulae which allow one to continue the asymptotic expressions for the eigenfunctions over
the whole spatial domain.
These expressions can then be used to finally construct the monodromy matrix.
Omitting the details for brevity (see Appendix~A.2 for details),}
\blue{we find the following expression for the trace of the monodromy matrix
}
%Nonetheless, after cumbersome calculations (see Appendix for the details), we find the following expression for the trace of the monodromy matrix
%Omitting all details for brevity (see Appendix for the detailed calculations), one finds
\vspace*{-0.2ex}
\begin{multline}
\kern-1em
\tr M =
%2 - 4\sin^2\left( \frac{S_1(\lambda)}{\epsilon}\right)\,\cosh^2\left[\frac{S_{2,\epsilon}(\lambda)}{\epsilon} \right],\quad 0<\lambda<1,
\\ \kern1em
  2 - 4\sin^2(S_1(\lambda)/\epsilon)\,\cosh^2[S_{2,\epsilon}(\lambda)/\epsilon],\quad 0<\lambda<1,
\label{e:trM_2}
\end{multline}
\ese
where $S_{2,\epsilon}(\lambda) = S_2(\lambda) + \epsilon\log 2$,
\bse
\vspace*{-0.6ex}
\begin{gather}
S_1(\lambda)=\int_{c}^{\pi-c}\sqrt{Q(x)}\,dx\,,
\\
S_2(\lambda)=\int_{-c}^{c}\sqrt{|Q(x)|}\,dx\,,
\end{gather}
\ese
and $c=\arccos(\sqrt{\lambda})$.

\blue{In the following, we conveniently refer to the half trace $\frac{1}{2}\tr M$, which, according to Eqs.~\eref{e:trM}, turns out to be bounded by the unit value,
i.e. $\frac{1}{2}\tr M\le1$, for all $\lambda\ge0$.}
%Note that $\tr M\le2$ for all $\lambda\ge0$, as follows from Eqs.~\eref{e:trM}.
In particular, Eq.~\eref{e:trM_1} shows that the range $1<\lambda<\infty$ forms one infinitely long band.
Conversely, Eq.~\eref{e:trM_2} shows that the range $0\le\lambda<1$ is divided into alternating bands and gaps
corresponding to the values of $\lambda$ for which $-1\le \frac{1}{2}\tr M \le 1$ (bands) or $\frac{1}{2} \tr M<-1$ (gaps), respectively.
The bands and gaps are separated by a sequence of band edges $\lambda_n$ for $n=0,1,2\dots$, which are the values of $\lambda$ such that $\frac{1}{2} \tr M = -1$.
%(corresponding to the values of $\lambda$ for which $-2\le \tr M\le2$ and $\tr M<-2$, respectively),
%separated by a sequence of band edges $\lambda_n$ for $n=0,1,2\dots$, which are the values of $\lambda$ for such that $\tr M = -2$.
Note that the value $\lambda=0$ is always part of the spectrum
(this can be seen by noting that for $k=0$ the scattering problem decouples in the variables $v$ and $\tilde v$,
and can be solved exactly to get $\frac{1}{2} \tr M=1$, which coincides with the limiting value of Eq.~\eref{e:trM_2} as $k\to0$).
%and can be solved exactly to get $\tr M=2$, which coincides with the limiting value of Eq.~\eref{e:trM_2} as $k\to0$).

\blue{
As an example, in Fig.~\ref{f:trM} {(left panel)} we show the dependence of $\frac{1}{2}\tr M$ on the spectral parameter $\lambda$ for $\epsilon=0.15$.
Note that, since $\tr M$ exhibits exponentially large oscillations, to capture the whole behavior in a single plot we plot the quantity $f(\tr M/2)$ instead of $\frac{1}{2}\tr M$ itself, as in~\cite{OsborneBergamasco},
with $f(\Delta)$ defined as $f(\Delta) = \Delta$ for $|\Delta|\le1$ and $f(\Delta) = \mathrm{sign}(\Delta)\,(1+\log_{10}|\Delta|)$ for $|\Delta|>1$.
In such figure we compare the above WKB asymptotic expressions for $\frac{1}{2}\tr M$ in Eq.~\eref{e:trM} with
the values obtained from direct numerical integration of Eq.~\eref{e:schrodinger}.
As can be seen from the figure, the difference between the asymptotic expressions and the numerical values is negligible for our purposes.
}
%As an example, in Fig.~\ref{f:trM} {(left panel)} we show the dependence of $\tr M$ on the spectral parameter $\lambda$ for $\epsilon=0.15$, where the above WKB asymptotic expressions for $\tr M$ in Eq.~\eref{e:trM} are compared with the values obtained from direct numerical integration of Eq.~\eref{e:schrodinger}.
%As can be seen from the figure, the difference between the asymptotic expressions and the numerical values is negligible for our purposes.
%Note that, since $\tr M$ exhibits exponentially large oscillations, to capture the whole behavior in a single plot we plot the quantity $f(\tr M/2)$ instead of $\tr M$ itself, as in~\cite{OsborneBergamasco}, with $f(\Delta)$ defined as $f(\Delta) = \Delta$ for $|\Delta|\le1$ and $f(\Delta) = \mathrm{sign}(\Delta)\,(1+\log_{10}|\Delta|)$ for $|\Delta|>1$.

%%%%%%%%%%%%%%%%%%%%%%%%%%%%%%%%%%%%%%%%%%%%%%%%%%%%%%%%%%%%%%%%%%%%%%%%%%%%%%%%%%%%%%%%%%%%%
\section{\noexpand\blue{III.~ Effective solitons}}

%%%%%%%%%%%%%%%%%%%%%%%%%%%%%%%%%%%%%%%%%%%%%%%%%%%%%%%%%%%%%%%%%%%%%%%%%%%%%%%%%%%%%%%%%%%%%
\subsection{III.1~ {Asymptotic properties of the spectrum}}

Equation~\eref{e:trM_2} implies that, for $0\le\lambda<1$, each band is clustered around a maximum of $\frac{1}{2}\tr M$, at which $\frac{1}{2}\tr M(\lambda) = 1$,
and that the $n$-th maximum $\lambda = z_n$ is given by the solution of
\vspace*{-0.2ex}
\[
S_1(z_n)= n\pi\epsilon\,.
\label{e:mndef}
\]
Since $\lambda = k^2$, all {spectral} bands of the scattering problem~\eref{e:schrodinger} correspond to a symmetric pair of nonlinear excitations of the NLS Eq.~\eref{nls}
except for the first {spectral} band, centered around $k=0$, which generates just one nonlinear excitation.
Hence, the number $N_e$ of nonlinear excitations of the problem is related to
the total number $N_b$ of {spectral} bands in the range $0\le\lambda<1$ by $N_e=2N_b-1$.
Since $S_1(1) = 2$, Eq.~\eref{e:mndef} then yields $N_b = \lfloor2/(\pi\epsilon)\rfloor+1$ and therefore
$N_e=2\lfloor2/(\pi\epsilon)\rfloor{{}+{}}1$.
%
% Sitai: I think plus one is correct, because there is always at least one excitation.
%
It was conjectured in \cite{trillovaliani,fatometrillo} that these excitations display soliton-like behavior.
For a fixed value of $\epsilon$, however, only some of the excitations in the problem resemble the dark solitons of the defocusing NLS equation.
Note however that counting the number of excitations from direct numerical simulations of the NLS equation presents two major challenges.
The first one is that not all excitations might be identifiable in the output.
The second one is that it is highly nontrivial to distinguish which ones among all visible excitations are of solitonic or non-solitonic type.
%
%In order to distinguish soliton-like excitations from non-solitonic ones, one must look at the the relative width $W_n$ of the $n$-th band,
%defined as the ratio between the band width $w_n$ and the sum of the band width and gap width $g_n$.
In order to distinguish soliton-like excitations from non-solitonic ones, one must look at the the relative width $W_n=w_n/(w_n+g_n)$ of the $n$-th band,
where $w_n$ ($g_n$) is the width of $n$-th band (adjacent gap).
By expanding the expression for $\tr M$ in Eq.~\eref{e:trM_2}, we find
\blue{(see Appendix~A.3 for details)}
\[
\!\! W_n=\frac{2}{\pi}\exp(-S_2(z_n)/\epsilon)+O(\epsilon \e^{-S_2(z_n)/\epsilon})\,.
\label{e:Wn}
\]

%----figure 2
\begin{figure}[t!]
\centering
\hglue-0.4em%
\includegraphics[scale=0.315,trim={0.15cm 0cm 0cm 0cm},clip]{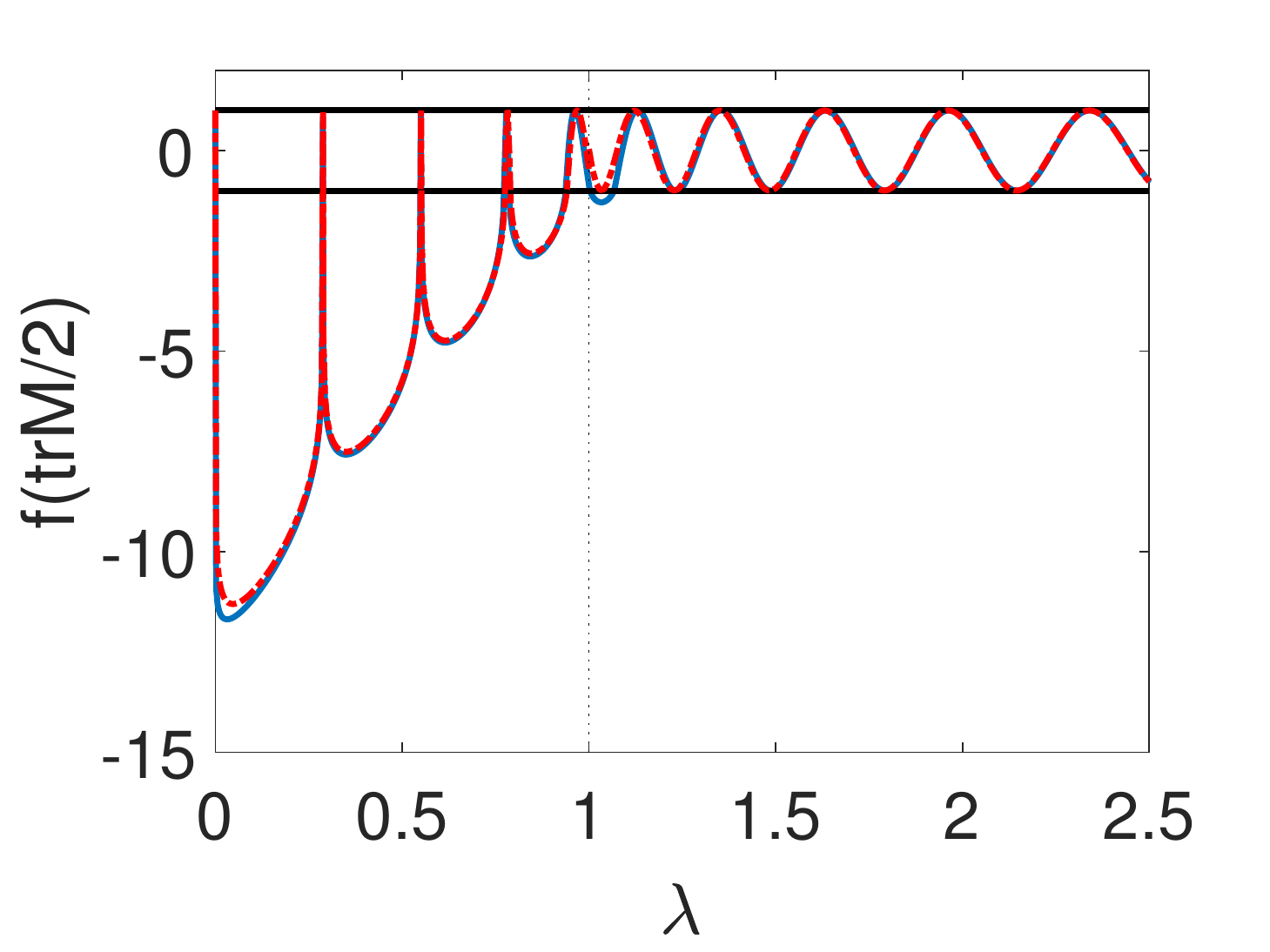}\kern0em%
\raise-0.1ex\hbox{\includegraphics[scale=0.315,trim={0cm 0cm 0cm 0cm},clip]{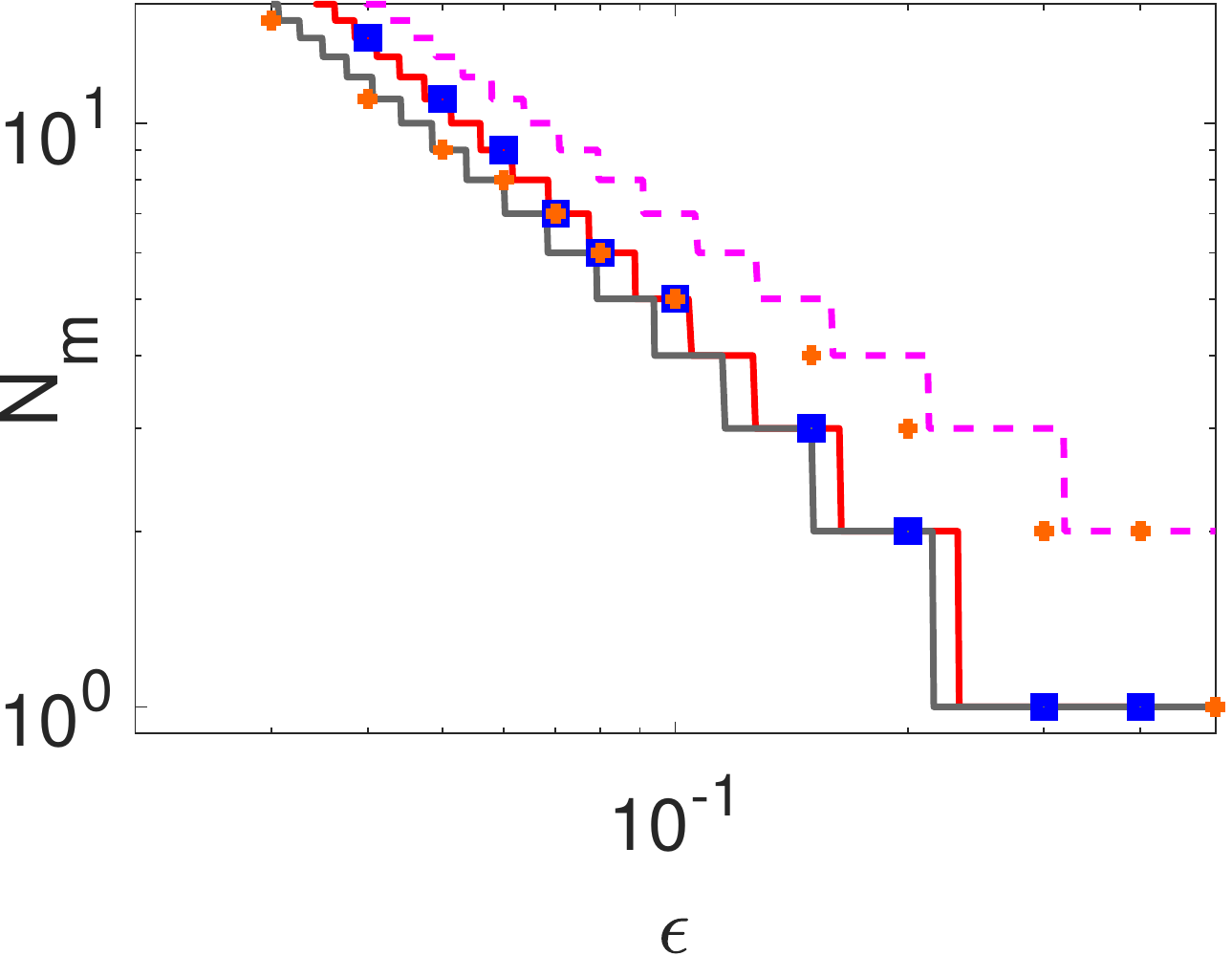}}%
%\includegraphics[scale=0.615,trim={0.15cm 0cm 0cm 0cm},clip]{trM2}\kern0em%
%\raise-0.1ex\hbox{\includegraphics[scale=0.615,trim={0cm 0cm 0cm 0cm},clip]{excitationnum}}%
\kern-\smallskipamount
\caption{
%(Color online) NO LONGER NEEDED
    Left panel: Half trace of the monodromy matrix $\frac{1}{2}\tr M$ as a function of the eigenvalue $\lambda$ with $\epsilon=0.15$, comparing
    the asymptotic expressions~\eref{e:trM} (dashed red) with the results from numerical integration of the scattering problem~\eref{e:schrodinger} (solid blue).
    Right panel: the number {$N_m$ of solitonic excitations} as a function of $\epsilon$.
    Red curve: full WKB prediction, Eq.~\eref{e:Nsoliton}.
    Gray curve: linear approximation, Eq.~\eref{e:Nsolitonslinear}.
%   Black curve: quadratic approximation of the number of solitons Eq.~\eref{e:Nsolitonsquadratic}.
    {Blue squares: the value obtained from numerical integration of Eq.~\eqref{e:zs}.
    Orange stars: same from direct numerical integration of Eq.~\eqref{nls}.
    Also shown {for comparison (magenta curve)} is the total number of excitations~$N_e$.
}
}
\label{f:trM}
\end{figure}

The proper solitonic limit of each excitation is obtained when the relative bandwidth $W_n$ tends to zero. Of course, since the bandwidths are always greater than zero,
one never has true solitons in the periodic problem.
(This is at variance, for instance, with what happens when one considers the dark potential $q(x,t=0)=\tanh x$ on the infinite line, which is reflectionless for integer $1/\epsilon$, containing a number $N_s=2/ \epsilon-1$ of discrete eigenvalues that correspond to true solitons, as shown in~\cite{Fratax08}.)\,
Nonetheless, excitations associated with very narrow relative bandwidths ($W_n \ll1$)
become closer and closer approximations of the dark solitons of the defocusing NLS equation.
Correspondingly, {given} a fixed threshold $\kappa\ll1$, similarly to \cite{physd2016deng,prl2016deng}
we define a nonlinear excitation of the periodic problem to be an \textit{effective soliton} if the band to which it is associated
is such that $W_n<\kappa$. Using Eq.~\eref{e:Wn} and solving this inequality, we then see that the solitonic bands are confined to the
range $[0,\lambda_s)$,
%Correspondingly, the value $\lambda_n = m_{n,\mathrm{approx}}$, i.e.,
%\[
%\lambda_n = 2n\epsilon\,,\qquad
%\epsilon\to0\,,
%\label{e:lambda_n}
%\]
%is a discrete eigenvalue with $\lambda_n\in[\,0,\lambda_s)$,
where $\lambda_s$ is defined by
\[
S_2(\lambda_s) = \epsilon\log[2/(\pi\kappa)].
\label{e:lambdasdef}
\]

Denoting by $N_m$ the total number of maxima in $[0,\lambda_s)$, {we have that $N_m$ is also the number of \textit{distinct soliton pairs}
(in which any two symmetric gray solitons are counted as one, like the lone black soliton). In light of the above discussion, from Eq.~\eref{e:mndef} we then have}
\vspace*{-0.6ex}
\[
{N_m = \lfloor S_1(\lambda_s)/(\pi\epsilon)+1\rfloor\,.}
%N_s = 2\big\lfloor S_1(\lambda_s)/(\pi\epsilon)\big\rfloor+1\,.
\label{e:Nsoliton}
\]
{The total number of solitons present in the problem is then simply $N_s = 2N_m-1$.}
As shown in Fig.~\ref{f:trM} {(right panel)}, the above WKB estimate for the total number of solitons is in excellent agreement
with results from numerical integration of Eq.~\eref{e:zs} and also in good agreement with direct numerical simulations of Eq.~\eref{nls}
(again with the caveat regarding the difficulty of counting numerically the number of excitations in case of small dispersion).

We can also derive {\em fully explicit} approximations for the number of solitons as a function of $\epsilon$
by considering either a linear or quadratic expansion of $S_1(\lambda)$ near $\lambda=0$ and a linear expansion of $S_2(\lambda)$ near $\lambda=1$
\blue{(see Appendix A.3 for details)}, thus obtaining
\bse
\label{e:Nsolitonsapprox}
\begin{gather}
{N_{m,\mathrm{linear}}} = \bigg\lfloor\frac{1}{2\epsilon}-\frac{1}{\pi}\log\frac{2}{\pi\kappa}+1\bigg\rfloor\,,
\label{e:Nsolitonslinear}
\\
\begin{aligned}
{N_{m,\mathrm{quadratic}}} = \bigg\lfloor \frac{4\sqrt{2} + 1}{8\sqrt{2}\epsilon} -
  \frac{2\sqrt{2} + 1}{2\sqrt{2}\pi}\log\frac{2}{\pi\kappa}+1\bigg\rfloor\,,
\label{e:Nsolitonsquadratic}
\end{aligned}
\end{gather}
\ese
respectively.
{Corresponding approximations for $N_s$ follow accordingly.}
The values {of $N_{m,\mathrm{linear}}$} are shown in Fig.~\ref{f:trM};
{those of $N_{m,\mathrm{quadratic}}$ are almost indistinguishable from the ``exact'' count given by Eq.~\eqref{e:Nsoliton}.}
Note that {in both cases} the leading order term is independent of the specific value chosen for the threshold $\kappa$.

%Notice that both $N$ and $N_\mathrm{approx}$ are decreasing functions of $\epsilon$,
%implying that smaller value of $\epsilon$ corresponds to more effective solitons.

%-------------------- figure 3
\begin{figure}[b!]
%\kern-2\smallskipamount
\centering
\hglue-0.5em
  \includegraphics[scale=0.325,trim={0.1cm 0.2cm 1cm 0cm},clip]{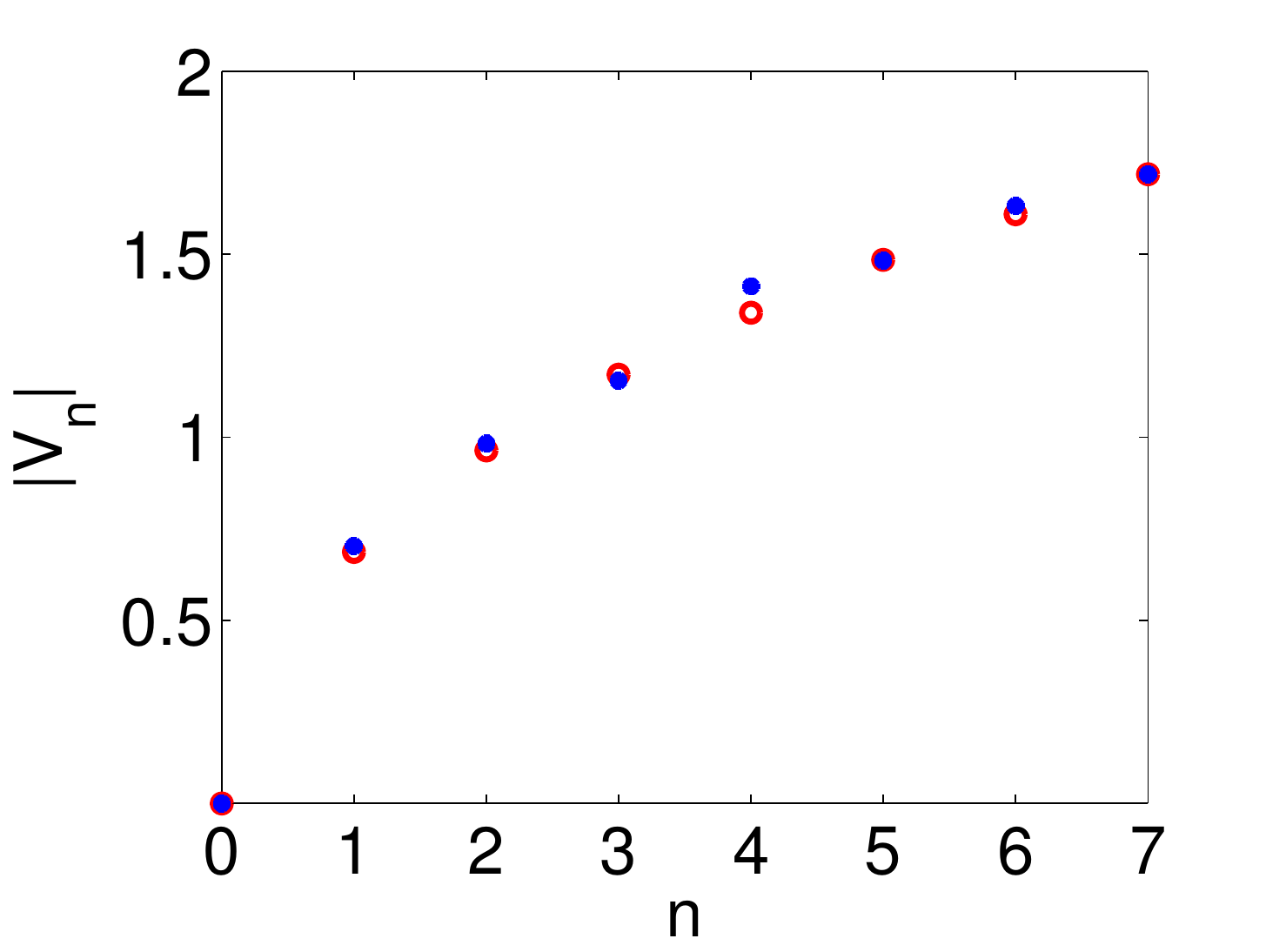}%
  \includegraphics[scale=0.325,trim={0.1cm 0.2cm 1cm 0cm},clip]{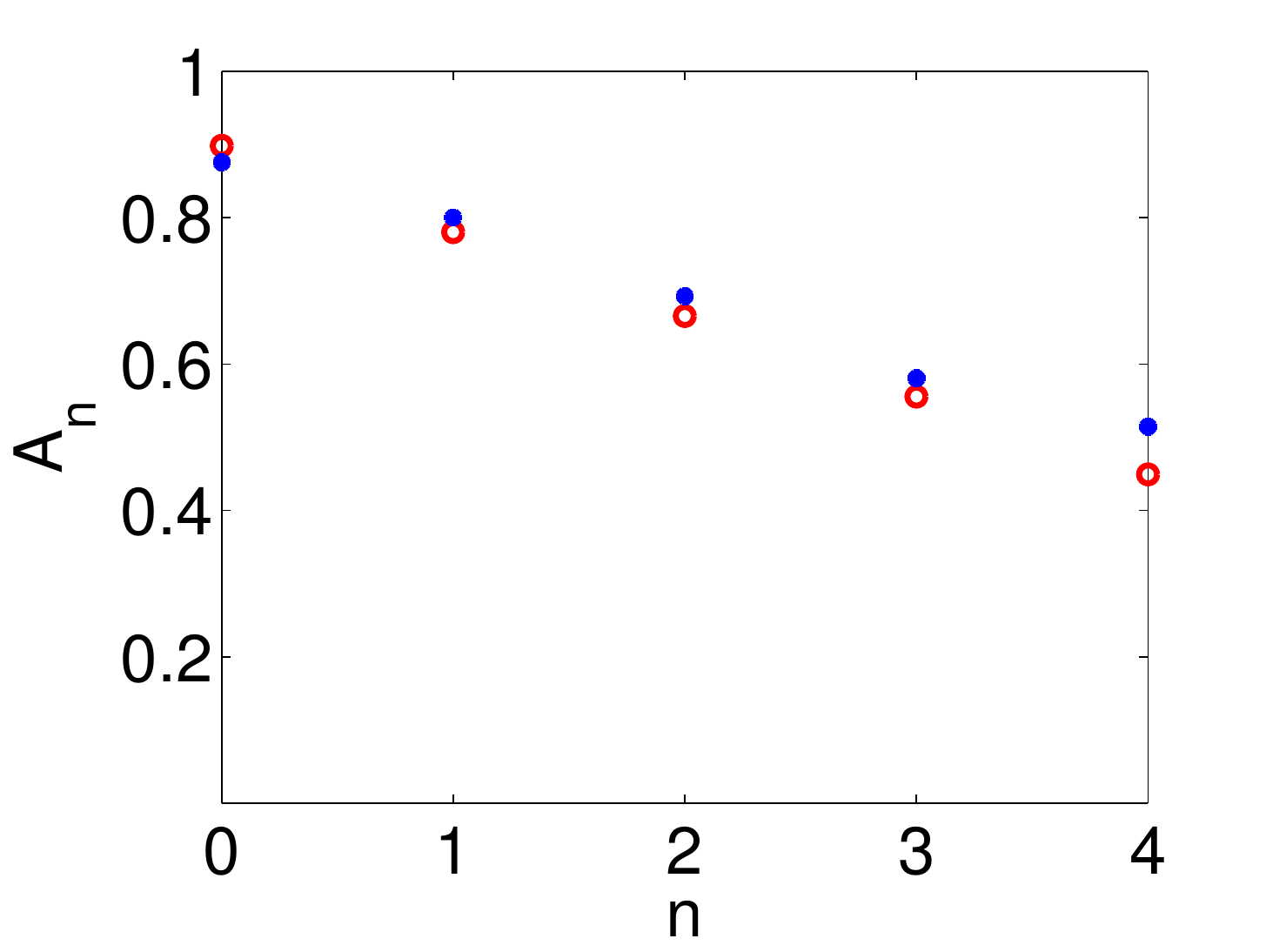}
\caption{
%(Color online)
  Absolute {value of} the soliton velocity $|V_n|$ {(left panel)} and amplitude $A_n$ {(right panel)} for the soliton
  {set associated to a given index} $n$.
  Red circles: asymptotic values; % Eq.~\eref{e:va},
  Blue dots: direct numerical simulations of Eqs.~\eref{nls}-\eref{ic}. Here $\epsilon=0.06$,
  {resulting in $N_s = 15$ and $q_\infty = 0.898$.}}
\label{f:kn}
\end{figure}

\smallskip
%%%%%%%%%%%%%%%%%%%%%%%%%%%%%%%%%%%%%%%%%%%%%%%%%%%%%%%%%%%%%%%%%%%%%%%%%%%%%%%%%%%%%%%%%%%%%
\subsection{III.2~ Amplitudes and velocities of the effective solitons}

Recall that the single dark-soliton solution of the defocusing NLS equation associated with a discrete eigenvalue $k = k_o$ is such that
\vspace*{-0.2ex}
\begin{multline}
\kern-1em
|q(x,t)|^2 = k_o^2
\\\textstyle
  + (q_\infty^2 - k_o^2)\tanh^2[\sqrt{q_\infty^2-k_o^2}(x - x_o - 2k_o t)]\,,
\label{e:darksoliton}
\end{multline}
where $q_\infty = \lim_{x\to\pm\infty}|q(x,t)|$ is the background amplitude.
Also recall that, to each solitonic band centered at {$\lambda=z_n\ne0$},
with $z_n$ defined by Eq.~\eref{e:mndef},
there corresponds a pair of dark solitons with discrete eigenvalues $k_{\pm n} = \pm \sqrt{z_n}$.
The band centered at ${z_0}=0$, instead, corresponds to a single black soliton.
We then obtain the velocity and the amplitude of the $n$-th {set} of solitons as
$V_n = \pm 2\sqrt{z_n}$ and $A_n = q_\infty^2-z_n$, respectively, {$n=0,1,2,\dots$}
Finally, recall that, in the infinite line problem, the continuous spectrum is $k\in(-\infty,-q_\infty)\cup(q_\infty,\infty)$.
Identifying the beginning of the continuous spectrum with the first non-solitonic band,
we therefore have $q_\infty = \sqrt{z_{N_m+1}}$.
%Recall the discrete eigenvalue $\lambda_n$ is given explicitly by Eq.~\eref{e:lambda_n}.
%Finally, we obtain asymptotic formulas for the $n$-th soliton as $\epsilon\to0$,
%\[
%V_n = 2\,\sign(n)\sqrt{2 |n| \epsilon}\,,\qquad
%A_n = (2|N_s+1|-2|n|)\epsilon\,,
%\label{e:va}
%\]
%where $-N\le n\le N$.
%As one can see, {$|V_n|$} and $A_n$ decreases and increases as $\epsilon\to0$, respectively.
Figure~\ref{f:kn} shows a comparison between the above WKB estimates for the soliton velocities and amplitudes
and the values obtained from direct numerical simulations of the NLS equation.

%\smallskip
%%%%%%%%%%%%%%%%%%%%%%%%%%%%%%%%%%%%%%%%%%%%%%%%%%%%%%%%%%%%%%%%%%%%%%%%%%%%%%%%%%%%%%%%%%%%%%
%\paragraph{Explanation of multiple wave mixing.}
%
%After obtaining the analytical description of the dispersionless limit of the defocusing NLS equation with cosine IC,
%we here explain the hydrodynamic instability,
%observed in~\cite{trillovaliani,fatometrillo}.
%The IC contains several effective solitons with soliton number given by Eq.~\eref{e:Nsoliton} or Eq.~\eref{e:Nsolitonsapprox}.
%As shown before, the number of solitons increases as $\epsilon\to0$,
%which coincides with experiments,
%i.e., as $\epsilon$ decreases, more peaks were observed.
%Moreover, based on WKB, the amplitude of each soliton increases as $\epsilon\to0$.
%This also explains why the soliton-like peaks became steeper in experiments
%as $\epsilon$ decreases.
%
%{Discuss the factor of $2$ from the total number of solitons,
%  theory vs numerical simulations.}
%

%--------------------- figure 4
\begin{figure}[t!]
\centering
\hglue-0.2em%
\includegraphics[scale=0.32]{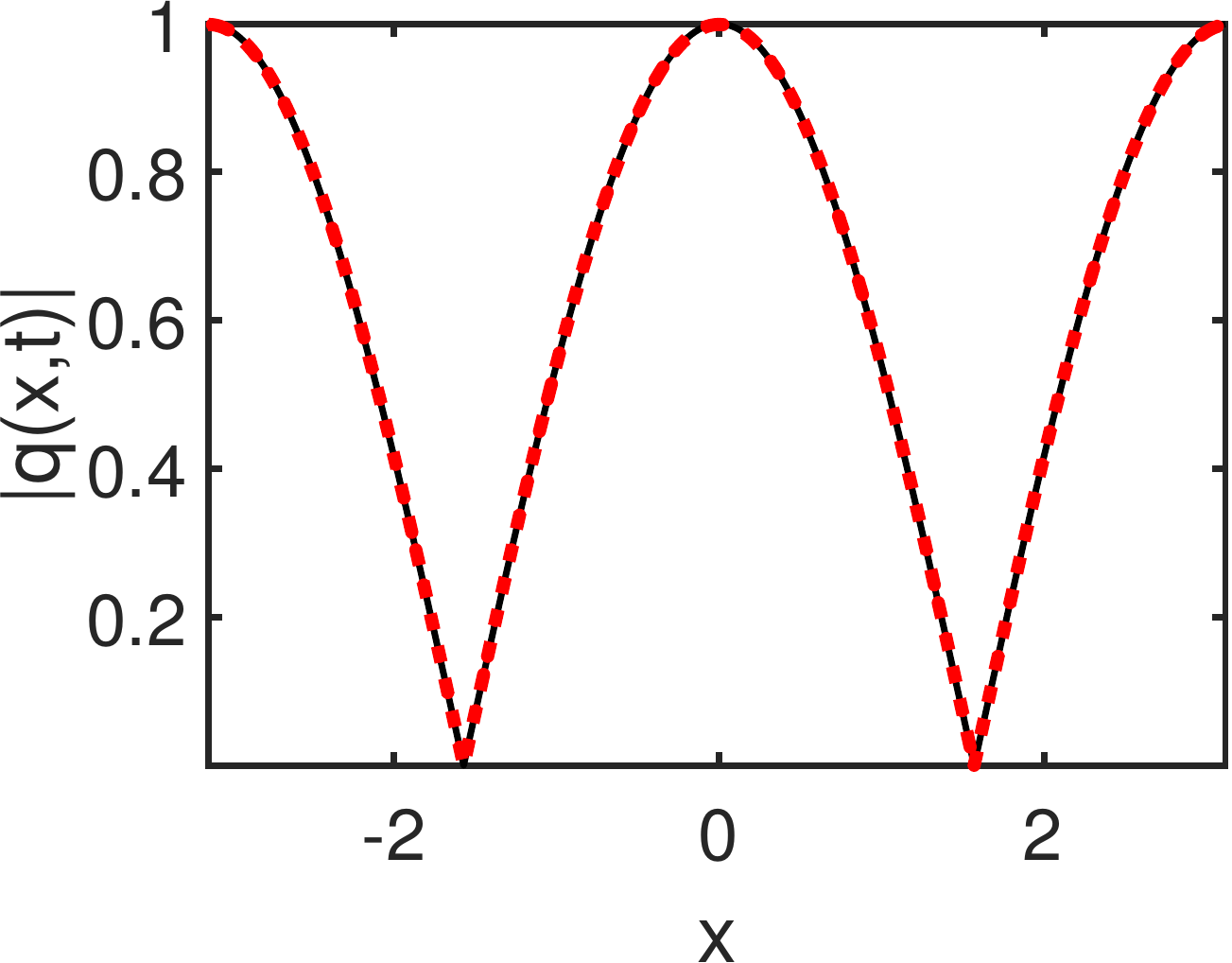}\,\,
\includegraphics[scale=0.32]{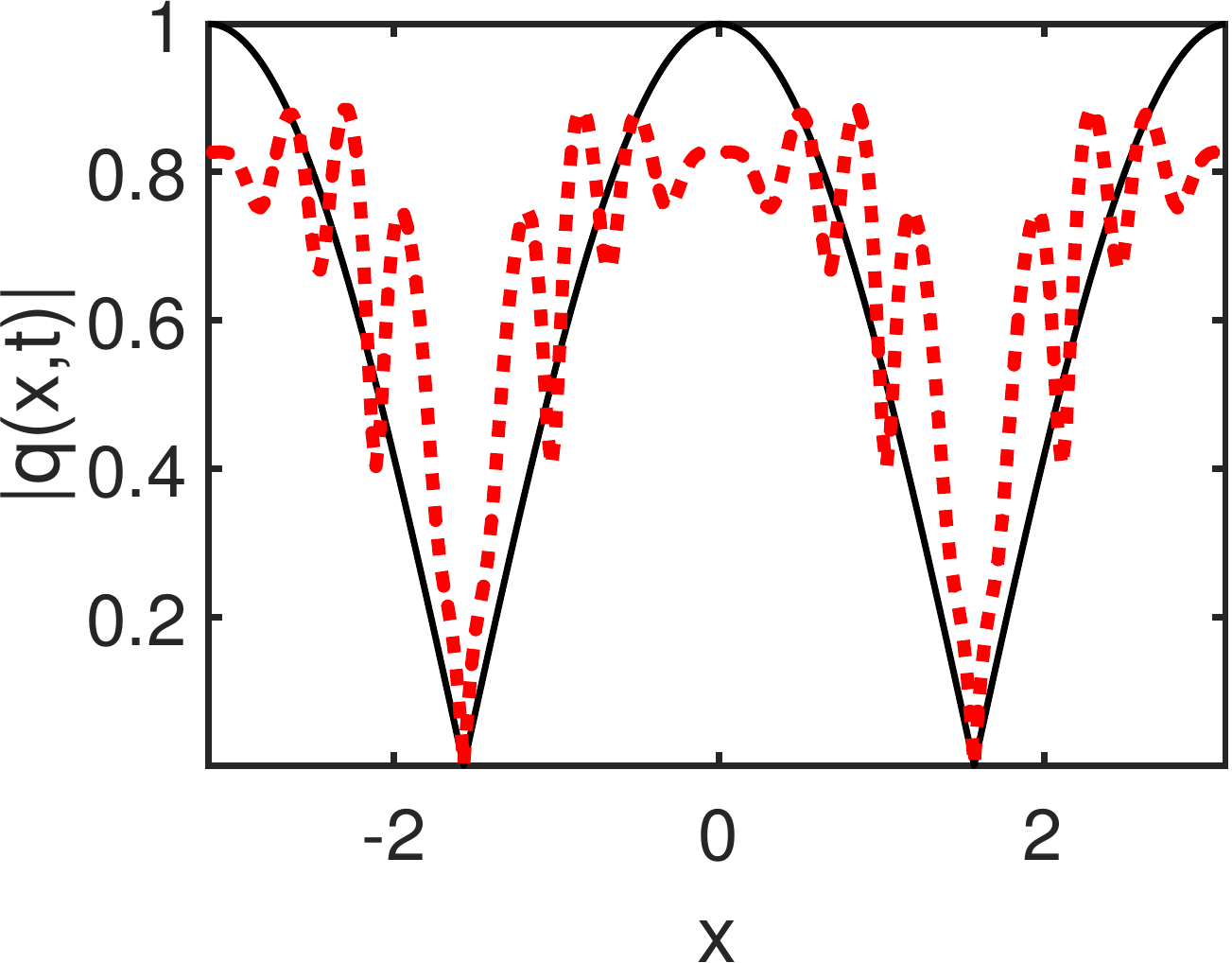}
\caption{
%(Color online)
Near-recurrence of initial condition for $\epsilon=0.3$ (left panel) and $\epsilon= 0.08$ (right panel),
comparing the initial value (\ref{ic}) (solid black) and $q(x,t)$ at the near-recurrence time (dashed red).
%Left panel: Recurrence of initial conditions for $\epsilon=0.3$, comparing the initial value (\ref{ic}) (solid black) and $q(x,t)$ at the near-recurrence time (dashed red curve). Right panel: Same for $\epsilon= 0.08$.
}
\label{f:recurrence}
\end{figure}
%\kern\bigskipamount
%\hglue-0.2em%

%%%%%%%%%%%%%%%%%%%%%%%%%%%%%%%%%%%%%%%%%%%%%%%%%%%%%%%%%%%%%%%%%%%%%%%%%%%%%%%%%%%%%%%%%%%%%
\section{\noexpand\blue{IV.~ Recurrence of initial conditions}}

Similarly to the KdV equation, the evolution ruled by the defocusing NLS equation is expected to nearly {recover} the cosine initial value
when the solitons simultaneously return to their initial location after traveling an integer number of periods.

\begin{figure}[t!]
\includegraphics[scale=0.32]{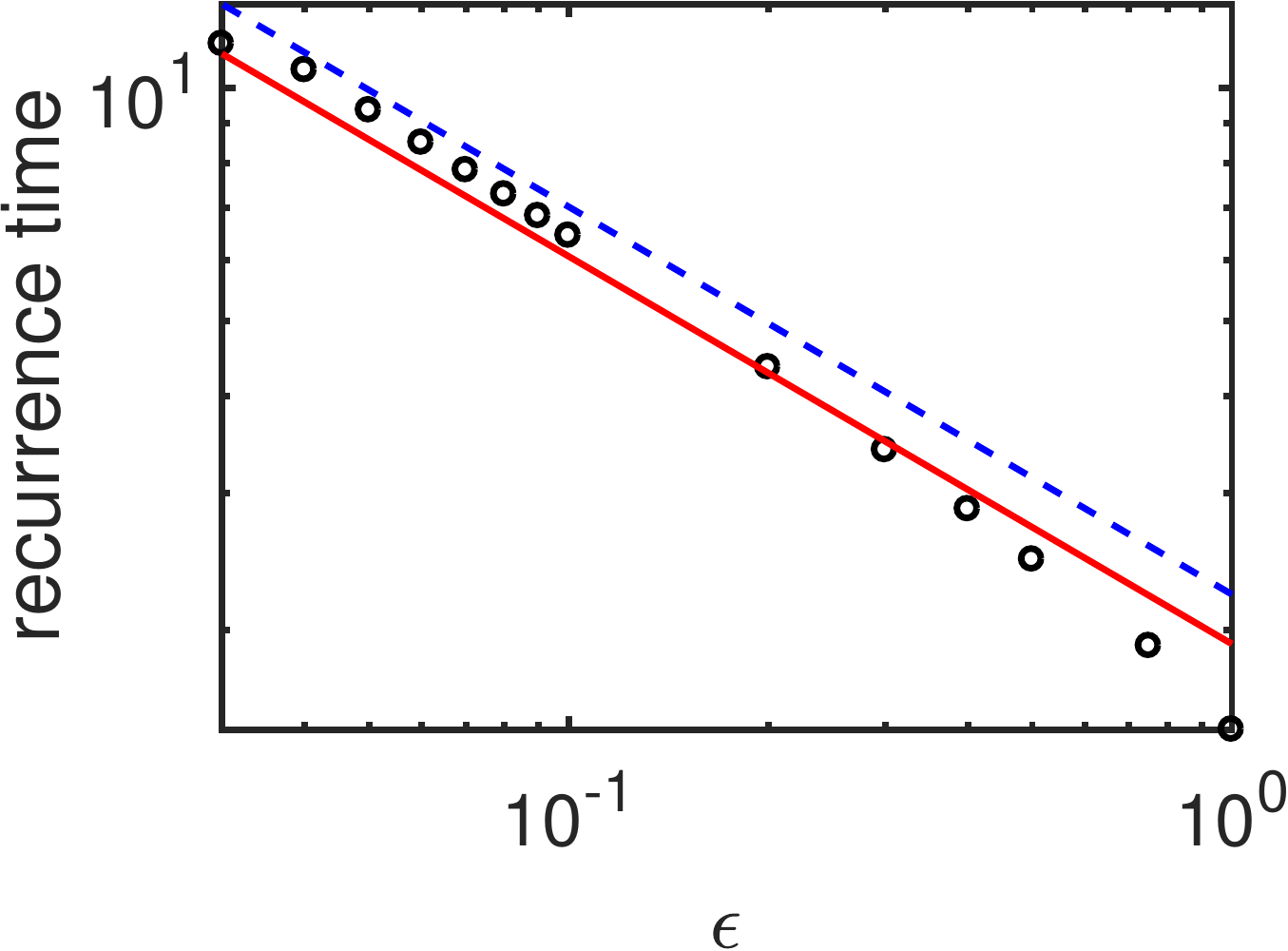}%
\includegraphics[scale=0.32]{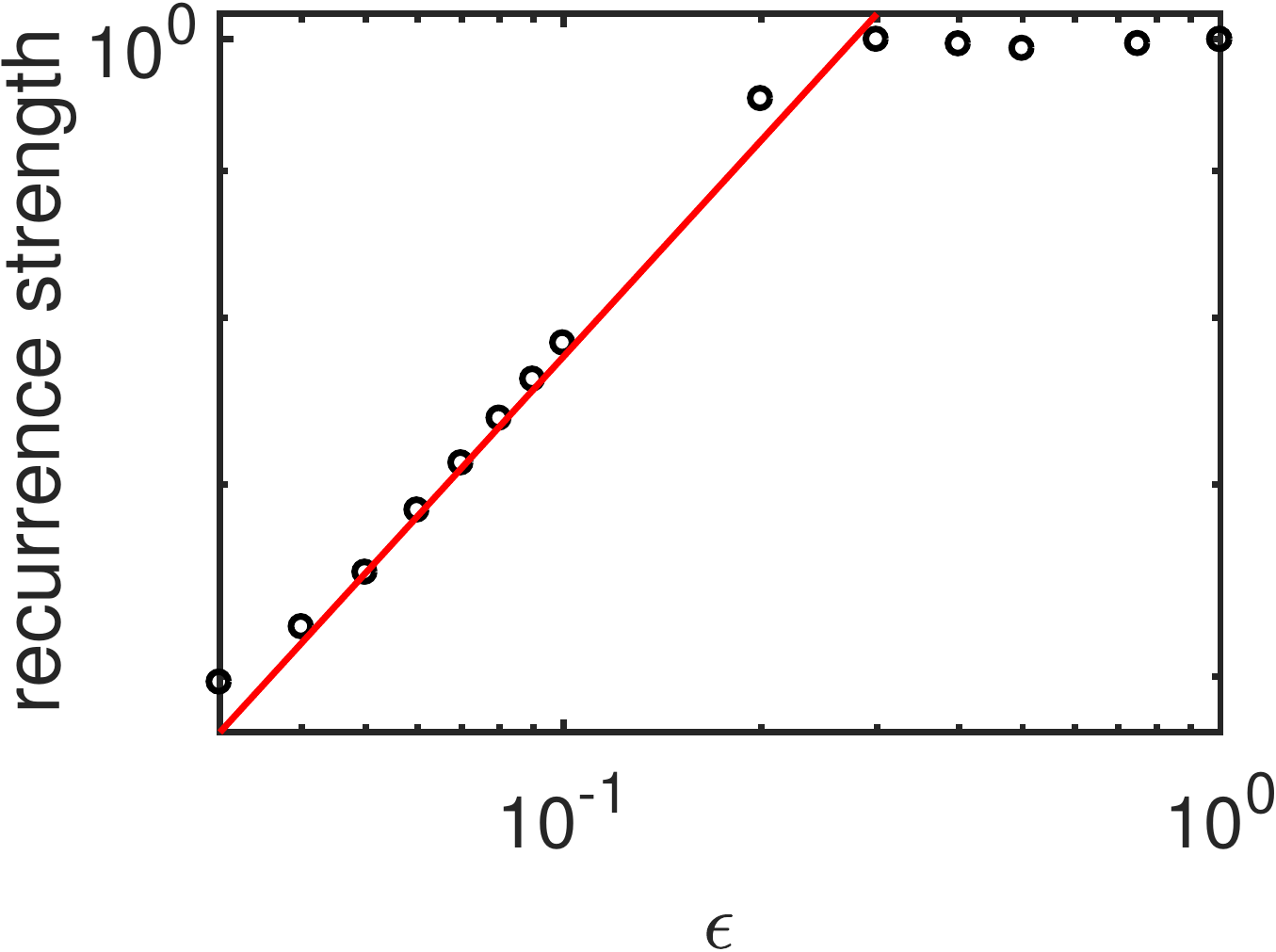}
\caption{
%(Color online)
Left panel: recurrence time as a function of $\epsilon$, as computed from direct numerical simulations of Eqs.~\eref{nls}-\eref{ic} (black circles),
along with a linear fit of the numerical data (red line) and the analytical prediction $t_\mathrm{recur} = t_{R1}$ (dashed blue line).
Right panel: Figure of merit of recurrence strength [Eq. (\ref{strength})], as computed from direct numerical simulations (black circles), and its linear fit for $\epsilon<0.3$ (red line).}
\label{f:recurrence2}
\end{figure}

Figure~\ref{f:recurrence} shows the solution at the (numerically determined) recurrence time for $\epsilon = 0.3$ and $\epsilon = 0.08$.
The numerically determined dependence of the recurrence time on $\epsilon$ is also shown in Fig.~\ref{f:recurrence2}.
Next we show how one can use the above WKB results to obtain an estimate of the recurrence time, \blue{where we neglect the small shifts
due to soliton collisions.}

%We also show, however, that there are significant differences between NLS and KdV recurrences.

\blue{
Since our purpose is also to study the differences between recurrences in the KdV and the NLS equations,
let us briefly consider first the case of the KdV, which we write as
\[
u_t + 6 u u_x + \epsilon^2 u_{xxx} = 0\,.
\label{e:kdv}
\]
An estimate for the recurrence time in Eq.~\eqref{e:kdv} can be obtained from the WKB analysis developed in Refs.~\cite{physd2016deng,prl2016deng}.
Adopting the same notation of such references, we recall that, in this case, the solitonic bands correspond to nonlinear excitations that are close to bright KdV solitons of the form
\begin{equation}
u_n(x,t)=u_0 + A_n ~{\rm sech}^2 \big[\sqrt{A_n/2}( x - V_nt )/\epsilon\big],
\end{equation}
where $A_n$ is the amplitude of the $n$-th soliton from the background level $u_0$, and $V_n=6u_0 + 2A_n$ is the corresponding velocity.
The amplitude can be calculated as \cite{OsborneBergamasco,physd2016deng}
$A_n=2(\lambda_\mathrm{ref}-\lambda_n)$ for $n=1,2,\ldots N$, where $N$ is the number of the solitonic bands and $\lambda_\mathrm{ref}=-u_0$ corresponds to the $(N+1)$-th band, or first non-solitonic band.
According to the analysis in Ref.~\cite{physd2016deng} and adopting a linear approximation for the eigenvalues, we find that the amplitudes and hence the velocities scales linearly
with the soliton order $n$, with increment $\Delta A=A_{n}-A_{n+1}=2\sqrt{2} \epsilon$ and hence $\Delta V=V_{n}-V_{n+1}=4\sqrt{2} \epsilon$.
Clearly, this means that after a time $T_{\mathrm{KdV}}$ such that $\Delta V\,T_{\mathrm{KdV}}=2\pi$,
all the solitons recur having traveled a multiple of the period (recall that $2 \pi$ is the period of the cosine,  and note that, obviously, the contribution to the velocity that is common to all solitons does not affect the recurrence time).
From this condition we then immediately find
\vspace*{-0.6ex}
\begin{equation}
\label{e:Tkdv}
T_{\mathrm{KdV}} = \frac{\pi}{2\sqrt{2}\,\epsilon}.
\end{equation}
Remarkably, the estimate in Eq.~\eqref{e:Tkdv} coincides with an earlier one by Toda \cite{toda1,toda2}
(once one performs the trivial rescalings of the spatial and temporal variables so as to obtain the same form of the KdV equation).
We note, however, that Toda's estimate was obtained using a quadratic (parabolic) approximation for the cosine potential, which amounts to approximating the eigenvalues with those of an harmonic oscillator.
\blue{Importantly, numerical results (not shown) based on the integration of the KdV equation confirm the scaling of the recurrence time with $\varepsilon^{-1}$ from Eq.~\eqref{e:Tkdv},
while the recurrence times from the numerics turn out to be slightly overestimated by the formula, possibly due to the soliton-soliton interactions, which are neglected in both our and Toda's approaches. }
}

Returning back to the defocusing NLS equation, the time needed for the $n$-th soliton pair to travel a distance equal to a whole period is $T_n = 2\pi/|V_n|$, with $V_n=\pm 2\sqrt{z_n}$ as before.
By employing again
%Performing
a linear expansion of $S_1(\lambda)$ around $\lambda=0$ (see Eqs.~(\ref{e:s1Taylor}) in Appendix),
we obtain the explicit expression $V_n=\pm2\sqrt{2n\epsilon}$, which in turn yields
\begin{equation}
\blue{T_n = \frac{\pi}{\sqrt{2n\epsilon}}\,.}
\end{equation}
This \blue{result} shows that the key difference between the NLS and KdV equations is that, for the latter, $V_n$ is linearly proportional to the soliton index $n$ (in the WKB limit),
whereas the above expression shows that for the NLS solitons $V_n$ depends on the square root of~$n$.
This difference is reflected in the travel time $T_n$.
Therefore, while in the KdV equation all solitons simultaneously return to their initial position at the recurrence time,
this is not true in the NLS equation, and the situation is more complicated in this case.

More precisely, neglecting the interaction-induced position shifts, for any fixed integer $m>0$, all soliton pairs with indices $n=m,4m,9m\dots,l^2m,\dots$ (with $l$ any {fixed} positive integer)
will return to their initial position, which is also the position of the stationary black soliton, at integer multiples of the recurrence time $t_{Rm}=\pi/\sqrt{2m\epsilon}$.
The value of $t_{R1}$ as a function of $\epsilon$, shown in Fig.~\ref{f:recurrence2}, is in good agreement with the numerically determined recurrence times.
Importantly, note that the recurrence time for the NLS equation scales like $\epsilon^{-1/2}$, whereas the recurrence time for the KdV equation scales like $\epsilon^{-1}$
%Importantly, note that the recurrence time for the NLS equation scales like $1/\sqrt{\epsilon}$, whereas the recurrence time for the KdV equation scales like $1/\epsilon$.

% former paragraph on KdV
%\blue{EXPAND ELSEWHERE} (specifically, for the KdV equation normalized as in \cite{physd2016deng}, a linear approximation of the soliton eigenvalues
%yields an estimate for the difference between the velocities of successive solitons as $\Delta V = 4\sqrt2\,\epsilon$,
%which in turn implies that all solitons recur at $t_{R,\mathrm{KdV}} = \pi/(2\sqrt{2}\,\epsilon)$ \cite{toda1,toda2}.

In order to further characterise the degree of the recurrence, we introduce the following figure of merit
\begin{equation}
R = 1 - \||q(x,t)|-|\cos x|\|/\|\cos x\|\,,
\label{strength} \end{equation}
where $\|f\|^2 = \int_0^{2\pi}|f(x)|^2\,dx$
{quantifies the energy contained in a periodic signal.} (Note that $\|\cos x\|^2 = \pi$.)
\blue{The value of $R$ is a measure of the strength of the recurrence.}
A perfect recurrence {corresponds to} $R=1$. The numerically determined dependence of $R$ on $\epsilon$ is reported in Fig.~\ref{f:recurrence2},
showing a strong deterioration for decreasing $\epsilon$, whereas near-perfect recurrence is achieved for relatively large values of $\epsilon$.
The latter fact is straightforwardly explained from Eq.~\eref{e:Nsoliton}, which implies that, for $\epsilon>0.163$,
there is only one soliton pair besides the {stationary} black soliton. In this regime, the initial condition is recovered almost exactly at multiples of $t_R=\pi/(\sqrt{2\epsilon})$,
as also illustrated in Fig.~\ref{f:recurrence} (left panel), consistently with {earlier} experiments \cite{Mamyshev94,Mamyshev94a} (se also \cite{trillovaliani}).
%The recurrence strength decreases as $\epsilon \to 0$, while near-perfect recurrence is achieved for larger values of $\epsilon$ \cite{Mamyshev94}.
%This result can be explained in a straightforward way. Indeed, the results of Eq.~\eref{e:Nsoliton} imply that, for $\epsilon>0.163$,
%there is only one soliton pair besides the stationary black soliton. Therefore, in this case one can expect the initial condition to be recovered almost exactly at multiples of $t_R=\pi/(\sqrt{2\epsilon})$,
%This situation is illustrated in Fig.~\ref{f:recurrence} (left panel). However, for $\epsilon<0.163$, multiple soliton pairs arise, and the recurrence becomes progressively worse as $\epsilon$ decreases, as illustrated in the right panel in Fig.~\ref{f:recurrence}.
However, for $\epsilon<0.163$, multiple soliton pairs come into play, and the recurrence becomes progressively worse for decreasing $\epsilon$, as illustrated for $\epsilon=0.08$ in Fig.~\ref{f:recurrence} (right panel).
Importantly, a similar calculation for the KdV equation also shows a recurrence degradation with decreasing $\epsilon$.
However, the scaling law {in the two models} is different.
{Specifically,} $R$ scales proportionally to $\epsilon$ for the KdV, and as $\epsilon^{1/4}$ for the NLS (see Fig.~\ref{f:recurrence}, right panel).
%{CORRECT ? consistent with initial report for KdV?}
% ST: OLD TEXT NOT CLEAR to ME,
% GB: I'm not sure what you mean.  Sitai's recurrence report shows the KdV recurrence strength scales like 1/epsilon.
%
%Importantly, it should be noted that the recurrence strength in the KdV equation also decreases with $\epsilon$.
%For the KdV equation, however, the recurrence strength is proportional to $\epsilon$.
%For the NLS equation, instead, a fit of the numerical data (see Fig.~\ref{f:recurrence2}) indicates a scaling law proportional to $\epsilon^{-1/4}$.
No analytical results are available in either model which explain this dependence.

%-----figure5

\section{\noexpand\blue{V.~ Conclusions}}

In summary, we have shown that, in systems governed by the weakly dispersing defocusing NLS equation, the fission of solitons from a periodic wave and their recurrence can be described in full analytical fashion.
The result demonstrates that the phenomenon of near-recurrence discovered by ZK is observable in other integrable systems, though important differences arise due to the different scaling of soliton velocities with their order.
{We believe that our} approach {can be easily} extended to {other} integrable systems
{(e.g., such as the Benjamin-Ono equation)}
in order to {produce} a critical assessment on the general degree of universality of the recurrence phenomenon originally discussed by ZK for the KdV equation.
We {also} expect our results to be {experimentally} verifiable in fiber optics {(in the context of} mFWM dynamics) and possibly in other settings (Bose-Einstein condensates, spin waves, oscillator {chains}).

\blue{The results of this work also open up the interesting question of whether a similar approach can also be used to study the small dispersion limit of the
focusing NLS equation with periodic boundary conditions.
Of course the focusing case is expected to be more challenging than the defocusing one.
From a mathematical point of view, this is because the corresponding spectral problem is no longer self-adjoint,
which means that the spectral bands are not restriced to the real axis,
and which puts into question whether one can effectively make use of the WKB method.
Physically, the problem is also expected to give rise to more complex phenomena because of the presence of modulational instability,
which becomes more and more severe in the dispersionless limit (i.e., as $\epsilon\to0$).
}

\vskip\bigskipamount

%\smallskip
%%%%%%%%%%%%%%%%%%%%%%%%%%%%%%%%%%%%%%%%%%%%%%%%%%%%%%%%%%%%%%%%%%%%%%%%%%%%%%%%%%%%%%%%%%%%%
\textbf{Acknowledgment.}
We thank A. Armaroli for insightful discussions during the early stage of this work.
This work was supported in part by the National Science Foundation under grant numbers DMS-1614623 and DMS-1615524.
S.T.\ acknowledges support from the Italian Ministry of University and Research
(PRIN-2012BFNWZ2)

\makeatletter
\let\tru@int=\int
\def\int{\mathop{\textstyle\tru@int}\limits}
\def\gl{\mathrel{\mathpalette\overl@ss>}}
\def\lg{\mathrel{\mathpalette\overgr@at<}}
\def\d{\mathrm{d}}
\def\Natural{\mathbb{N}}
\def\Integer{\mathbb{Z}}
\def\Real{\mathbb{R}}
\def\Complex{\mathbb{C}}
\def\Re{\mathop{\rm Re}\nolimits}
\def\Im{\mathop{\rm Im}\nolimits}
\def\arg{\mathop{\rm arg}\nolimits}
\def\pvint{\mathop{\int\kern-0.94em-\kern0.2em}\limits}
\let\^=\hat
\let\==\bar
\let\@=\mathbf
\let\letrue=\le
\let\getrue=\ge
\let\le=\leqslant
\let\ge=\geqslant
\def\d{\mathrm{d}}
\def\e{\mathrm{e}}
\def\o{\mathrm{o}}
\def\Wr{\mathrm{Wr}}
\def\tr{\mathrm{tr}}
\def\Ai{\mathrm{Ai}}
\def\Bi{\mathrm{Bi}}
\def\sech{\mathrm{sech}}
\def\O#1{^{(#1)}}
\def\~#1{\tilde{#1}}
\makeatother

\def\be{\begin{equation}}
\def\ee{\end{equation}}
\def\bse{\begin{subequations}}
\def\ese{\end{subequations}}

\appendix
\section{APPENDIX}

Here we provide the details on the analytical calculation of the monodromy matrix and the relative bandwidths.

%%%%%%%%%%%%%%%%%%%%%%%%%%%%%%%%%%%%%%%%%%%%%%%%%%%%%%%%%%%%%%%%%%%%%%%%%%%%%%%%%%%%%%%%%%%%%
\subsection{\blue{A.1~ The WKB expansion and the range $\lambda>1$}}

\paragraph{The WKB expansion.}
%\kern-\medskipamount
%\noindent
Recall that the scattering problem is given by Eq.~\eref{e:schrodinger}.
We look for an asymptotic expansion for the solution $v$ of Eq.~\eref{e:schrodinger}
in the form
\vspace*{-1ex}
\be
v(x)=\left[ A(x)+O(\epsilon) \right]\,\e^{iS(x)/\epsilon}\,,\quad
\label{e:asymp}
\ee
as $\epsilon\to0$.
Substituting Eq.~\eref{e:asymp} into Eq.~\eref{e:schrodinger}, one obtains,
at the first two orders in the expansion,
the eikonal equation and the transport equation, i.e.,
\vspace*{-1ex}
\bse
\begin{gather}
S_x^2=Q(x),\\
2iS_x A_x + iS_{xx}A + \sin x\,A=0,
\end{gather}
\ese
respectively.
%where for brevity we denoted $A(x) = A_0(x)$.
These equations are readily integrated to obtain
\vspace*{-1ex}
\bse
\label{e:SandA}
\begin{gather}
S_\pm(x)=\pm\int\sqrt{Q(x)}\,dx\,,\\
A_\pm(x) = \sqrt{\sqrt{Q(x)}\mp i\cos x}\,\bigg/\sqrt[4]{Q(x)}\,,
\end{gather}
\ese
up to arbitrary additive and multiplicative constants, respectively,
where $Q(x) = \lambda - \cos^2 x$ was defined in Eq.~\eref{e:Qdef},
and $A_\pm$ corresponds to the plus/minus sign in $S_\pm(x)$, respectively.

Observe that,
for $\lambda>1$, $Q(x)>0$ for all $x\in R$.
Conversely, for $0<\lambda<1$ the range $x\in[-\pi,\pi]$ divides into three sub-domains:
\\[0.6ex]
\hbox to 2em{\hss(i)}~ $Q(x)<0$ for $x\in[-\pi,-c_2)\bigcup(-c_1,c_1)\bigcup(c_2,\pi]$,
\\[0.6ex]
\hbox to 2em{\hss(ii)}~ $Q(x)>0$ for $x\in(-c_2,-c_1)\bigcup(c_1,c_2)$,
\\[0.6ex]
\hbox to 2em{\hss(iii)}~ $Q(x)=0$ for $x=\pm c_1,\pm c_2$,
\\[0.6ex]
where $c_1=\arccos(\sqrt{\lambda})$
and $c_2=\pi- c_1$
(with $c = c_1$).
%This is illustrated in Fig.~\ref{f:Q(x)}.
We then need to study the WKB solutions in these two ranges of $\lambda$ separately.
%\begin{figure}[b!]
%\centerline{\includegraphics[width=0.465\textwidth]{Qx.pdf}}
%\caption
%{$Q(x)=k^2-\cos^2x$ as a function of x over $[-\pi,\pi]$.}
%\label{f:Q(x)}
%\end{figure}

\paragraph{Trace of the monodromy matrix for $\lambda>1$.}
%
%\kern-\medskipamount
%\nobreak\noindent
For any value of $\lambda$ in this range, $Q(x)$ is positive.
Two linearly independent solutions are given, in the WKB approximation, by
\vspace*{-1ex}
\be
\label{e:w1w2}
v_\pm(x)= A_\pm(x)\,\e^{iS_\pm(x)/\epsilon},
%w_2(x)= A_-\e^{-i\int\sqrt{Q(x)}\,dx/\epsilon},
\ee
with $A_\pm(x)$ and $S_\pm(x)$ given by Eq.~\eref{e:SandA}.
We can write a corresponding fundamental matrix solution of the first-order system associated with Eq.~\eref{e:schrodinger}
as a Wronskian:
\vspace*{-1ex}
\be
Y(x)=\mathrm{Wr}(v_-,v_+)\,.
\label{e:fundamental_matrix_solution}
\ee
%with $v_\pm$ given in Eq.~\eref{e:w1w2}.
Since $Q(x)\ne0$ in this case, such a fundamental matrix solution is valid over the whole range $x\in[-\pi,\pi]$.
%We first evaluate $W(x)$ at $x=-\pi$ and $x=\pi$.
%The value of this fundamental matrix solution
%at $x=-\pi$ and at $x=\pi$ is, respectively,
%\begin{subequations}
%\begin{align}
%&W(-\pi)= \begin{pmatrix}A_+(-\pi)& A_-(-\pi)\\
%\frac{i}{\epsilon}\sqrt{|Q(-\pi)|}A_+(-\pi)&-\frac{i}{\epsilon}\sqrt{|Q(-\pi)|}A_-(-\pi)\end{pmatrix},\\
%&W(\pi)= \begin{pmatrix}A_+(\pi)& A_-(\pi)\\
%\frac{i}{\epsilon}\sqrt{|Q(\pi)|}A_+(\pi)&-\frac{i}{\epsilon}\sqrt{|Q(\pi)|}A_-(\pi)\end{pmatrix},
%\end{align}
%\end{subequations}
%with $A_\pm$ given by~\eref{e:SandA}.
Therefore the mondoromy matrix can be obtained simply as
\vspace*{-1ex}
\begin{equation}
M = Y^{-1}(-\pi)Y(\pi)\,.
\label{e:monodromy}
\end{equation}
Straightforward calculations then show that the trace of $M$ is given by Eq.~\eref{e:trM_1}.
%namely
%\be
%\nonumber
%\tr M= 2\cos\bigg(\int_{-\pi}^{\pi}\sqrt{Q(x)}dx/\epsilon\bigg)\,.
%\ee
%Note that $|\tr M|\leq 2$, therefore all values of $\lambda$ in this range belong to one long stable band.

\begin{figure}[b!]
\medskip
\centerline{\includegraphics[width=0.465\textwidth]{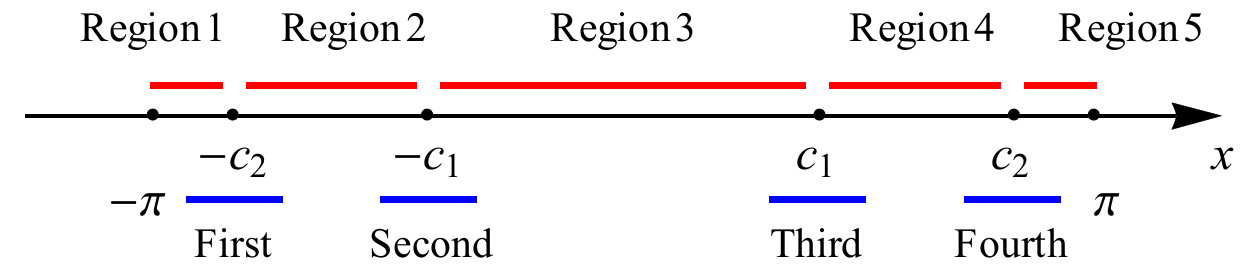}}
\caption
{The sub-regions of the fundamental domain $x\in[-\pi,\pi]$ for the WKB analysis in the range $0<\lambda<1$.}
\label{f:wkb_regions}
\end{figure}

%%%%%%%%%%%%%%%%%%%%%%%%%%%%%%%%%%%%%%%%%%%%%%%%%%%%%%%%%%%%%%%%%%%%%%%%%%%%%%%%%%%%%%%%%%%%%
\subsection{\blue{A.2~ Trace of the monodromy matrix for $0<\lambda<1$}}

\paragraph{WKB solutions for $0<\lambda<1$.}
%
%\kern-\medskipamount
%\noindent
For all values of $\lambda$ in this range, there are four turning points,
namely $\pm c_1,\pm c_2$.
Correspondingly we need to discuss the behavior of the eigenfunctions in the following nine sub-regions of the fundamental period $x\in[-\pi,\pi]$:
\vspace*{-1ex}
\begin{itemize}
\advance\itemsep-6pt
\item[(a)] Region~1, $x\in[-\pi, -c_2)$;
\item[(b)] First transition region, comprised by a neighborhood of $x=-c_2$;
\item[(c)] Region~2, $x\in(-c_2,-c_1)$;
\item[(d)] Second transition region, comprised by a neighborhood of $x=-c_1$;
\item[(e)] Region~3, $x\in(-c_1, c_1)$;
\item[(f)] Third transition region, comprised by a neighborhood of $x=c_1$;
\item[(g)] Region~4, $x\in(c_1, c_2)$;
\item[(h)] Fourth transition region, comprised by a neighborhood of $x=c_2$;
\item[(i)] Region~5, $x\in(c_2, \pi]$.
\end{itemize}

The various sub-regions are illustrated in Fig.~\ref{f:wkb_regions}.
Next, we discuss the WKB solution of the scattering problem in the various sub-regions.

\begingroup
\newcounter{region}
\setcounter{region}0
\long\def\item{\par\smallskip\stepcounter{region}(\alph{region})~~}

\item
  In region 1, $x\in[-\pi,-c_2)$, the WKB approximation for the general solution of the scattering problem~\eref{e:schrodinger} is
\vspace*{-1ex}
  \be
  v_1(x)=a_1^+v_{1+}(x)+a_1^-v_{1-}(x),
  \ee
  where
\vspace*{-1ex}
  \begin{equation}
  %\begin{aligned}
  v_{1\pm}(x)= A_\pm\exp(\mp\int_{-c_2}^x\sqrt{|Q(x)}|dx/\epsilon),\\
  %&w_2(x)= A_-\exp(\int_{c_1}^x\sqrt{|Q(x)}|dx/\epsilon),
  \label{e:solution_region1}
  %\end{aligned}
  \end{equation}
  and
  \be
  A_\pm=\big[\pm|Q(x)|^{1/2}-q\big]^{1/2}\big/|Q(x)|^{1/4}\,.
  \label{e:A}
  \ee

\item
  In the first transition region, $x\in(-c_2-\delta, -c_2+\delta)$ with $\delta>0$, Eq.~\eref{e:schrodinger} becomes
\vspace*{-1ex}
  \be
  \nonumber
  \epsilon^2 v_{xx}+(2k\sqrt{1-k^2}(x+c_2)-\epsilon\sqrt{1-k^2})v=0,
  \ee
  an asymptotic expansion for $v$ in the first transition region is given by
\vspace*{-1ex}
  \be
  \nonumber
  v_{1\to2}(x)=c_1^-\Ai[\xi(x)]+c_1^+\Bi[\xi(x)],
  \ee
  where
\vspace*{-1ex}
  \be
  \nonumber
  \xi(x)= -a^{1/3}(x+c_2-\epsilon/2k)/\epsilon^{2/3}, \quad
  a=2k\sqrt{1-k^2}.
  \ee

\item
  In region 2, $x\in(-c_2,-c_1)$,
  the WKB approximation to the general solution has two equivalent expressions
\vspace*{-1ex}
  \bse
  \begin{multline}
  v_2(x) = a_2^+A_+(x)\exp(i\int_{-c_2}^x\sqrt{|Q(x)}|dx/\epsilon)\\
    +a_2^-A_-(x)\exp(-i\int_{-c_2}^x\sqrt{|Q(x)|}dx/\epsilon),
  \label{e:region2expression1}
  \end{multline}
  \vspace*{-4ex}
  \begin{multline}
  v_2(x) = \bar{a}_2^+A_+(x)\exp(i\int_{-c_1}^x\sqrt{|Q(x)}|dx/\epsilon)\\
    +\bar{a}_2^-A_-(x)\exp(-i\int_{-c_1}^x\sqrt{|Q(x)|}dx/\epsilon),
  \label{e:region2expression2}
  \end{multline}
  \ese
  with $A_\pm$ given by Eq.~\eref{e:SandA}.

\item
  In the second transition region, i.e., $x\in(-c_1-\delta,-c_1+\delta)$, Eq.~\eref{e:schrodinger} becomes
\vspace*{-1ex}
  \be
  \nonumber
  \epsilon^2 v_{xx}-(2k\sqrt{1-k^2}(x+c_1)+\epsilon\sqrt{1-k^2})v=0\,.
  \ee
  An asymptotic expansion for $v$ in the second transition region is given by
\vspace*{-1ex}
  \be
  \nonumber
  v_{2\to3}(x)=c_2^-\Ai[\eta(x)]+c_2^+\Bi[\eta(x)],
  \ee
  where $\eta(x)= a^{1/3}(x+c_1+\epsilon/2k)/\epsilon^{2/3}$.

  \item
  In region 3, $x\in(-c_1,c_1)$, the general solution in the WKB approximation also has two equivalent expressions
\vspace*{-1ex}
  \bse
  \begin{multline}
  v_3(x) = a_3^+A_+(x)\exp(-\int_{-c_1}^x\sqrt{|Q(x)}|dx/\epsilon)\\
  +a_3^-A_-(x)\exp(\int_{-c_1}^x\sqrt{|Q(x)|}dx/\epsilon),
  \label{e:region3expression1}
  \end{multline}
  \vspace*{-4ex}
  \begin{multline}
  v_3(x) = \bar{a}_3^+A_+(x)\exp(-\int_{c_1}^x\sqrt{|Q(x)}|dx/\epsilon)\\
  +\bar{a}_3^-A_-(x)\exp(\int_{c_1}^x\sqrt{|Q(x)|}dx/\epsilon),
  \label{e:region3expression2}
  \end{multline}
  \ese
  where $A_\pm=\big[\mp|Q(x)|^{1/2}+q\big]^{1/2}\big/|Q(x)|^{1/4}$.

  \item
  In the third transition region,
  $x\in(c_1-\delta,c_1+\delta)$, Eq.~\eref{e:schrodinger} becomes
\vspace*{-1ex}
  \be
  \nonumber
  \epsilon^2 v_{xx}+(2k\sqrt{1-k^2}(x-c_1)+\epsilon\sqrt{1-k^2})v=0\,.
  \ee
  An asymptotic expansion for $v$ in the third transition region is given by
\vspace*{-1ex}
  \be
  \nonumber
  v_{3\to4}(x)=c_3^-\Ai[z(x)]+c_3^+\Bi[z(x)],
  \ee
  where $z(x)= -a^{1/3}(x-c_1+\epsilon/2k)/\epsilon^{2/3}$.

  \item
  In Region 4, $x\in(c_1,c_2)$, the WKB approximation to the general solution again has two equivalent expressions
\vspace*{-1ex}
  \bse
  \begin{multline}
  v_4(x) = a_4^+A_+(x)\exp(i\int_{c_1}^x\sqrt{|Q(x)}|dx/\epsilon)\\
  +a_4^-A_-(x)\exp(-i\int_{c_1}^x\sqrt{|Q(x)|}dx/\epsilon),
  \label{e:region4expression1}
  \end{multline}
  \vspace*{-4ex}
  \begin{multline}
  v_4(x) = \bar{a}_4^+A_+(x)\exp(i\int_{c_2}^x\sqrt{|Q(x)}|dx/\epsilon)\\
  +\bar{a}_4^-A_-(x)\exp(-i\int_{c_2}^x\sqrt{|Q(x)|}dx/\epsilon),
  \label{e:region4expression2}
  \end{multline}
  \ese
  with $A_\pm$ given by Eq.~\eref{e:SandA}.

\item
  In the fourth transition region,
  $x\in(c_2-\delta,c_2+\delta)$, Eq.~\eref{e:schrodinger} becomes
\vspace*{-1ex}
  \be
  \nonumber
  \epsilon^2 v_{xx}-(2k\sqrt{1-k^2}(x-c_2)-\epsilon\sqrt{1-k^2})v=0,
  \ee
  an asymptotic expansion for $v$ in the fourth transition region is given by
  \be
  \nonumber
  v_{4\to5}(x)=c_4^-\Ai[\gamma(x)]+c_4^+\Bi[\gamma(x)],
  \ee
  where $\gamma(x)= a^{1/3}(x-c_2-\epsilon/2k)/\epsilon^{2/3}$.

  \item
  In region 5, $x\in(c_2,\pi]$, the general solution in the WKB approximation is
\vspace*{-1ex}
  \begin{multline*}
  v_5(x) = a_5^+A_+(x)\exp(-\int_{c_2}^x\sqrt{|Q(x)}|dx/\epsilon)\\
  +a_5^-A_-(x)\exp(\int_{c_2}^x\sqrt{|Q(x)|}dx/\epsilon),
  \end{multline*}
  with $A_\pm$ given by Eq.~\eref{e:A}.

\endgroup

%%%%%%%%%%%%%%%%%%%%%%%%%%%%%%%%%%%%%%%%%%%%%%%%%%%%%%%%%%%%%%%%%%%%%%%%%%%%%%%%%%%%%%%%%%%%%
\paragraph{Connection formulae for $0<\lambda<1$.}
%
%\kern-\medskipamount
%\noindent
Having computed WKB expressions for the solution of the scattering problem in the various sub-domains for $0<\lambda<1$,
we now need to match the solutions across different sub-regions.
First we match $v_1(x)$ with $v_{1\to 2}(x)$. Explicitly,
\vspace*{-0.4ex}
\begin{gather*}
\begin{aligned}
&v_1(x)=\frac1{\sqrt[4]{a|x+c_2|}}(a_1^+\sqrt[4]{\lambda}\e^{\frac{2}{3}\sqrt{a}|x+c_2|^{3/2}/\epsilon}\\
&+a_1^-
\sqrt[4]{\lambda}\e^{-\frac{2}{3}\sqrt{a}|x+c_2|^{3/2}/\epsilon})(1+O(|x+c_2|)),\\
&x\to -c_2\,,
\end{aligned}
\\
\begin{aligned}
&v_{1\to 2}(x)=\frac{1}{\sqrt[4]{\pi^2\xi}}(\tfrac{1}{2} c_1^- \e^{-\frac{2}{3}\xi^{3/2}}+
c_1^+ \e^{\frac{2}{3}\xi^{3/2}})\\
&\times(1+O(1/\xi^{3/2})),\quad
\xi\to\infty.
\end{aligned}
\end{gather*}
Requiring that these two expansions match, we obtain the connection formula
\vspace*{-0.4ex}
\begin{equation*}
\begin{pmatrix}
     c_1^- \\  c_1^+
   \end{pmatrix} =
   C_1 \begin{pmatrix}
    a_1^-\\
    a_1^+
   \end{pmatrix},
\qquad
C_1 =    \frac{\sqrt[4]{\pi^2 \lambda}}{(a\epsilon)^{1/6}}
   \begin{pmatrix}
  2  &\   0 \\
  0  &\ 1
  \end{pmatrix}.
\end{equation*}

To match $v_2(x)$ with $v_{1\to 2}(x)$, note that
\begin{gather*}
\begin{aligned}
&v_2(x) = \frac1{\sqrt[4]{a(x+c_2)}}(a_2^+\sqrt[4]{\lambda}\e^{i\pi/4}\e^{\frac{2}{3}i\sqrt{a}(x+c_2)^{3/2}/\epsilon}\\
&+a_2^-\sqrt[4]{\lambda}\e^{-i\pi/4}\e^{-\frac{2}{3}i\sqrt{a}(x+c_2)^{3/2}/\epsilon})(1+O(x+c_2)),\\
& x\to -c_2\,,
\end{aligned}
\\
\begin{aligned}
&v_{1\to2}(x)  = \frac{1}{\sqrt[4]{\pi^2 |\xi|}}\big[ c_1^-\sin(\tfrac{2}{3}|\xi|^{3/2}+ \tfrac{\pi}{4})\\
&+ c_1^+\cos(\tfrac{2}{3}|\xi|^{3/2}+\tfrac{\pi}{4})\big](1+O(1/|\xi|^{3/2})),
\quad
\xi\to-\infty\,,
\end{aligned}
\end{gather*}
implying
\vspace*{-0.4ex}
\begin{equation*}
\begin{pmatrix}
     a_2^+ \\
     a_2^- \\
   \end{pmatrix}=
 C_2
    \begin{pmatrix}
   c_1^-\\
   c_1^+\\
   \end{pmatrix},
C_2=
  \frac{(a\epsilon)^{1/6}}{2\sqrt[4]{\pi^2 \lambda}}
   \begin{pmatrix}
   -i  &\  1\\
   i  &\  1\\
  \end{pmatrix}.
\end{equation*}

\begin{figure}[t!]
\kern2\medskipamount
\hglue-0.4em \includegraphics[scale=0.315,trim={0.1cm 0.2cm 1cm 0cm},clip]{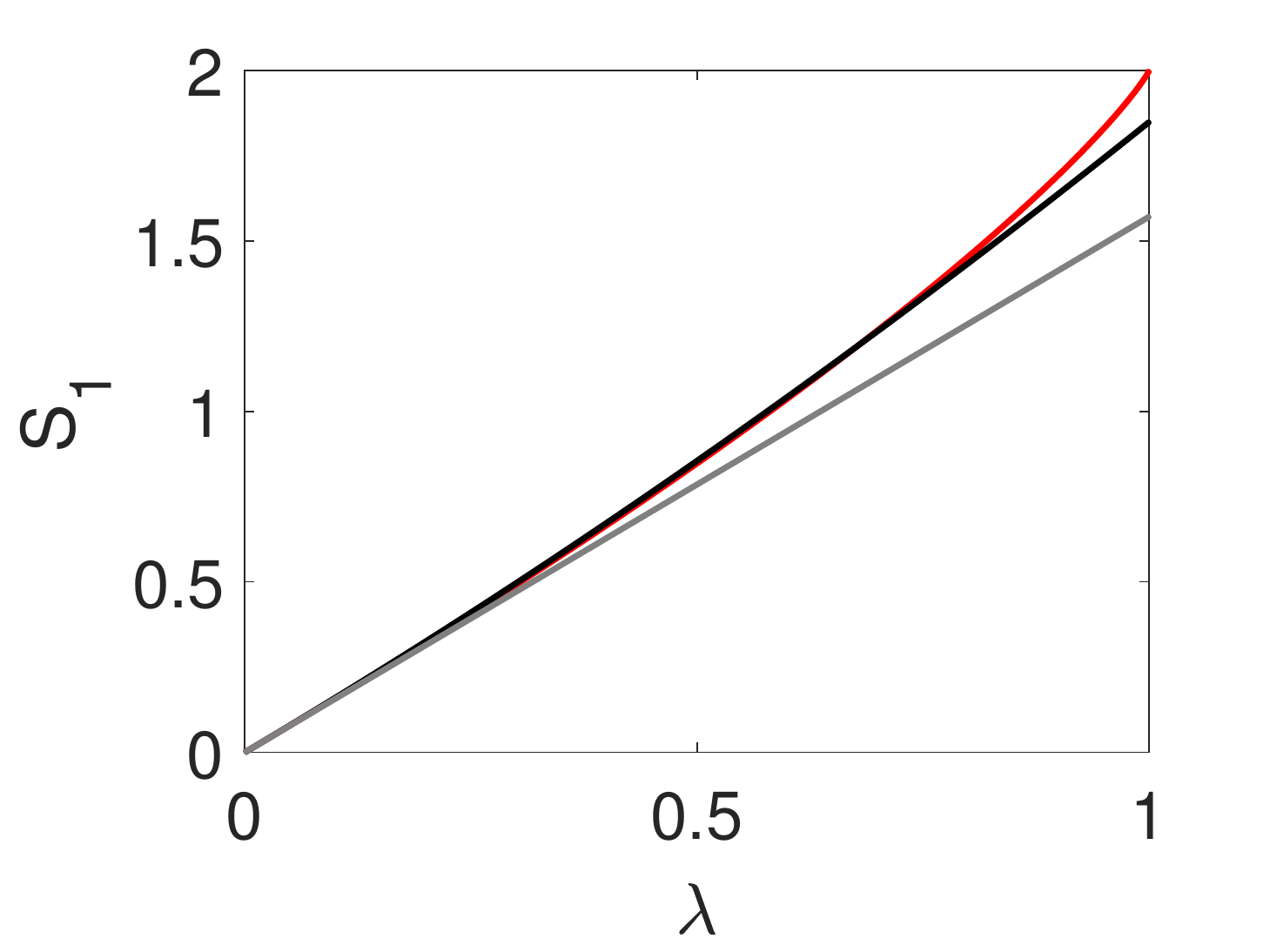}\kern1em
  \includegraphics[scale=0.315,trim={0.1cm 0.2cm 1cm 0cm},clip]{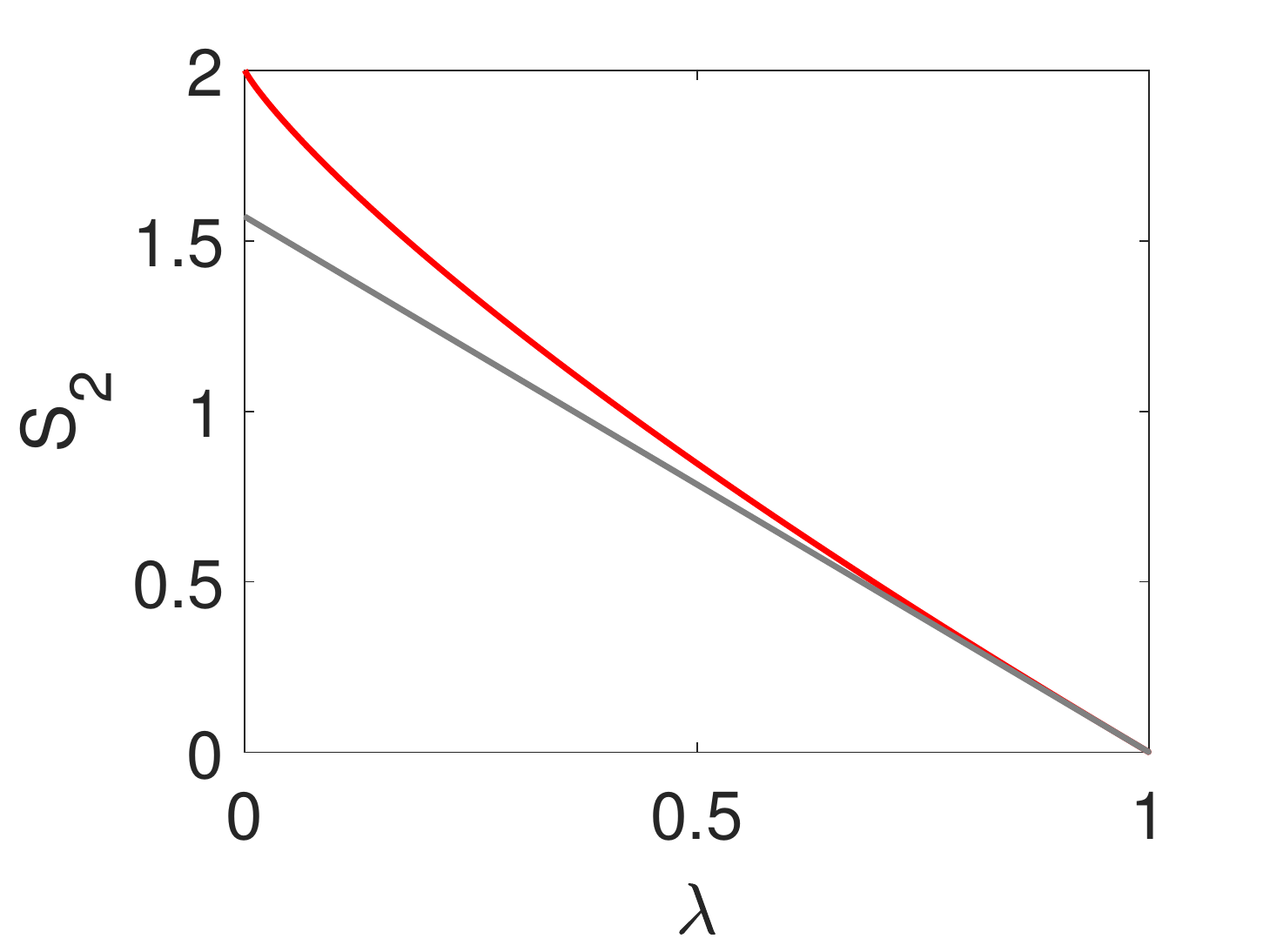}
  \caption{Left: $S_1(\lambda)$ as a function of $\lambda$ (red curve).
    The gray (black) curve shows the linear (quadratic) approximation of $S_1(\lambda)$ about $\lambda=-1$.
    Right: $S_2(\lambda)$ as a function of $\lambda$ (red curve).
    The gray curve shows the linear approximation of $S_2(\lambda)$ about $\lambda=1$.}
\label{f:S1S2}
\end{figure}

Note also that $\bar{a}_2^\pm$ in Eq.~\eref{e:region2expression2} relates to $a_2^\pm$ in Eq.~\eref{e:region2expression1} by
\vspace*{-1ex}
\begin{equation*}
\begin{pmatrix}
     \bar{a}_2^+ \\
     \bar{a}_2^- \\
   \end{pmatrix}=
    C_3
  \begin{pmatrix}
   a_2^+\\
   a_2^-\\
   \end{pmatrix}\,,\quad
C_3= \e^{i\sigma_3\, S_1(\lambda)/\epsilon},
\end{equation*}
with
\vspace*{-1ex}
\be
S_1(\lambda)=\int_{c_1}^{c_2}\sqrt{Q(x)}dx.
\label{e:S_1}
\ee
A plot of $S_1(\lambda)$ is shown in Fig.~\ref{f:S1S2}.

Next we match $v_2(x)$ with $v_{2\to3}(x)$. Note that
\vspace*{-1ex}
\begin{gather*}
\begin{aligned}
&v_2(x)  = \frac1{\sqrt[4]{a|x+c_1|}}(\bar{a}_2^-\sqrt[4]{\lambda}\e^{i\pi/4}\e^{\frac{2}{3}i\sqrt{a}|x+c_1|^{3/2}/\epsilon}\\
&+
\bar{a}_2^+\sqrt[4]{\lambda}\e^{-i\pi/4}\e^{-\frac{2}{3}i\sqrt{a}|x+c_1|^{3/2}/\epsilon})
(1+O(x+c_1)),\\
&x\to -c_1\,,
\end{aligned}
\\
\begin{aligned}
&v_{2\to3}(x)  = \frac{1}{\sqrt[4]{\pi^2 |\eta|}}\big[ c_2^-\sin(\tfrac{2}{3}|\eta|^{3/2}+ \tfrac{\pi}{4})\\
&+ c_2^+\cos(\tfrac{2}{3}|\eta|^{3/2}+\tfrac{\pi}{4})\big](1+O(1/|\eta|^{3/2})),
\quad
\eta\to -\infty\,.
\end{aligned}
\end{gather*}
Requiring that these two expansions match, we obtain the connection formula
\vspace*{-1ex}
\vspace*{-0.4ex}
\begin{equation*}
\begin{pmatrix}
     c_2^+ \\  c_2^-
   \end{pmatrix} =
   C_4 \begin{pmatrix}
    \bar{a}_2^+\\
    \bar{a}_2^-
   \end{pmatrix}\,,\quad
C_4 =    \frac{\sqrt[4]{\pi^2 \lambda}}{(a\epsilon)^{1/6}}
   \begin{pmatrix}
  1 &\  1 \\
  -i  &\  i
  \end{pmatrix}.
\end{equation*}

Next we match $v_3(x)$ with $v_{2\to3}(x)$. Note that
\vspace*{-0.4ex}
\begin{gather*}
\begin{aligned}
&v_{3}(x)=\frac{1}{\sqrt[4]{a(x+c_1)}}(a_3^-\sqrt[4]{\lambda}\e^{\frac{2}{3}\sqrt{a}(x+c_1)^{3/2}/\epsilon}\\
&+a_3^+\sqrt[4]{\lambda}\e^{-\frac{2}{3}\sqrt{a}(x+c_1)^{3/2}/\epsilon})(1+O(x+c_1)),\\
& x\to -c_1\,,
\end{aligned}
\\
\begin{aligned}
&v_{2\to3}(x)=\frac{1}{\sqrt[4]{\pi^2\eta}}(\tfrac{1}{2} c_2^-\e^{-\frac{2}{3}\eta^{3/2}}+c_2^+\e^{\frac{2}{3}\eta^{3/2}})\\
&\times(1+O(1/\eta^{3/2})),
\quad\eta\to\infty\,.
\end{aligned}
\end{gather*}
Matching these expansions, we obtain
\vspace*{-0.4ex}
\begin{equation*}
\begin{pmatrix}
     a_3^- \\
     a_3^+ \\
   \end{pmatrix}=
   C_5
   \begin{pmatrix}
    c_2^+\\
    c_2^-\\
   \end{pmatrix}\,,\qquad
C_5  = \frac{(a\epsilon)^{1/6}}{\sqrt[4]{\pi^2 \lambda}}
   \begin{pmatrix}
   1 &\  0\\
   0  &\  \frac{1}{2} \\
  \end{pmatrix}.
\end{equation*}

Note that $\bar{a}_3^\pm$ in Eq.~\eref{e:region3expression2} relates to $a_3^\pm$ in Eq.~\eref{e:region3expression1} by
\vspace*{-0.4ex}
\begin{equation*}
\begin{pmatrix}
     \bar{a}_3^- \\
     \bar{a}_3^+ \\
   \end{pmatrix}=
    C_6
  \begin{pmatrix}
   a_3^-\\
   a_3^+\\
   \end{pmatrix}\,,\qquad
C_6= \e^{\sigma_3 S_2(\lambda)/\epsilon}
\end{equation*}
with
\vspace*{-1ex}
\be
S_2(\lambda)=\int_{c_2}^{c_3}\sqrt{|Q(x)|}dx.
\label{e:S_2}
\ee
A plot of $S_2(\lambda)$ is shown in Fig.~\ref{f:S1S2}.

To match $v_3(x)$ with $v_{3\to 4}(x)$, note that
\vspace*{-0.4ex}
\begin{gather*}
\begin{aligned}
&v_3(x)=\frac1{\sqrt[4]{a|x-c_1|}}(\bar{a}_3^+\sqrt[4]{\lambda}\e^{\frac{2}{3}\sqrt{a}|x-c_1|^{3/2}/\epsilon}\\
&+\bar{a}_3^-
\sqrt[4]{\lambda}\e^{-\frac{2}{3}\sqrt{a}|x-c_1|^{3/2}/\epsilon})(1+O(|x-c_1|)),
\\
&x\to c_1\,,
\end{aligned}
\\
\begin{aligned}
&v_{3\to 4}(x) = \frac{1}{\sqrt[4]{\pi^2z}}(\tfrac{1}{2} c_3^- \e^{-\frac{2}{3}z^{3/2}}+
c_3^+ \e^{\frac{2}{3}z^{3/2}})\\
&\times(1+O(1/z^{3/2})),\quad z\to\infty.
\end{aligned}
\end{gather*}
Requiring that these two expansions match, we obtain the connection formula
\begin{eqnarray*}
   \begin{pmatrix}
     c_3^- \\  c_3^+
   \end{pmatrix} =
   C_1 \begin{pmatrix}
    \bar{a}_3^-\\
    \bar{a}_3^+
   \end{pmatrix}.
\qquad
%C_7 =    \frac{\sqrt{\pi k}}{(a\epsilon)^{1/6}}
%   \begin{pmatrix}
%  2  &\   0 \\
%  0  &\ 1
%  \end{pmatrix}.
\end{eqnarray*}

Next we match $v_{4}(x)$ with $v_{3\to 4}(x)$, note that
\vspace*{-0.4ex}
\begin{gather*}
\begin{aligned}
&v_4(x) = \frac1{\sqrt[4]{a(x-c_1)}}(a_4^+\sqrt[4]{\lambda}\e^{-i\pi/4}\e^{\frac{2}{3}i\sqrt{a}(x-c_1)|^{3/2}/\epsilon}\\
&+a_4^-\sqrt[4]{\lambda}\e^{i\pi/4}\e^{-\frac{2}{3}i\sqrt{a}(x-c_1)^{3/2}/\epsilon})(1+O(x-c_1)),\\
&x\to c_1\,,
\end{aligned}
\\
\begin{aligned}
&v_{3\to4}(x) = \frac{1}{\sqrt[4]{\pi^2 |z|}}\big[c_3^-\sin(\tfrac{2}{3}|z|^{3/2}+ \tfrac{\pi}{4})\\
&+c_3^+\cos(\tfrac{2}{3}|z|^{3/2}+\tfrac{\pi}{4})\big](1+O(1/|z|^{3/2})),
\quad z\to-\infty\,.
\end{aligned}
\end{gather*}
Matching these expansions, we obtain
\vspace*{-0.4ex}
\begin{equation*}
\begin{pmatrix}
     a_4^+ \\
     a_4^- \\
   \end{pmatrix}=
 C_7
    \begin{pmatrix}
   c_3^-\\
   c_3^+\\
   \end{pmatrix}\,,\quad
C_7=
  \frac{(a\epsilon)^{1/6}}{2\sqrt[4]{\pi^2 \lambda}}
   \begin{pmatrix}
   1  &\  i\\
   1  &\  -i\\
  \end{pmatrix}.
\end{equation*}

Note also that $\bar{a}_4^\pm$ in Eq.~\eref{e:region4expression2} relates to $a_4^\pm$ in Eq.~\eref{e:region4expression1} by
\vspace*{-0.4ex}
\begin{eqnarray*}
   \begin{pmatrix}
     \bar{a}_4^+ \\
     \bar{a}_4^- \\
   \end{pmatrix}=
    C_3
  \begin{pmatrix}
   a_4^+\\
   a_4^-\\
   \end{pmatrix}.
%   C_9=
%     \begin{pmatrix}
%  \e^{iS_2(k)/\epsilon} &\  0\\
%   0  &\  \e^{-iS_2(k)/\epsilon}\\
%  \end{pmatrix}.
 \end{eqnarray*}
Then we match $v_{4}(x)$ with $v_{4\to5}(x)$. Note that
\begin{gather*}
\begin{aligned}
&v_{4}(x) = \frac1{\sqrt[4]{a|x-c_2|}}(\bar{a}_4^-\sqrt[4]{\lambda}\e^{-i\pi/4}\e^{\frac{2}{3}i\sqrt{a}|x-c_2|^{3/2}/\epsilon}\\
&+\bar{a}_4^+\sqrt[4]{\lambda}\e^{i\pi/4}\e^{-\frac{2}{3}i\sqrt{a}|x-c_2|^{3/2}/\epsilon})(1+O(x-c_2)),\\
&x\to c_2\,,
\end{aligned}
\\
\begin{aligned}
&v_{4\to5}(x) = \frac{1}{\sqrt[4]{\pi^2 |\gamma|}}\big[ c_4^-\sin(\tfrac{2}{3}|\gamma|^{3/2}+ \tfrac{\pi}{4})\\
&+ c_4^+\cos(\tfrac{2}{3}|\gamma|^{3/2}+\tfrac{\pi}{4})\big](1+O(1/|\gamma|^{3/2})),
\quad \gamma\to-\infty\,.
\end{aligned}
\end{gather*}
Requiring that these two expansions match, we obtain the connection formula
\vspace*{-0.4ex}
\begin{equation*}
\begin{pmatrix}
     c_4^+ \\  c_4^-
   \end{pmatrix} =
   C_8 \begin{pmatrix}
    \bar{a}_4^+\\
    \bar{a}_4^-
   \end{pmatrix}\,,\quad
C_8 =  \frac{\sqrt[4]{\pi^2 \lambda}}{(a\epsilon)^{1/6}}
   \begin{pmatrix}
  i &\  -i \\
  1  &\  1
  \end{pmatrix}.
\end{equation*}

Finally we math $v_{5}(x)$ with $v_{4\to5}(x)$. Note that
\vspace*{-0.4ex}
\begin{gather*}
\begin{aligned}
&v_{5}(x)=\frac{1}{\sqrt[4]{a(x-c_2)}}(a_5^-\sqrt[4]{\lambda}\e^{\frac{2}{3}\sqrt{a}(x-c_2)^{3/2}/\epsilon}\\
&+a_5^+\sqrt[4]{\lambda}\e^{-\frac{2}{3}\sqrt{a}(x-c_2)^{3/2}/\epsilon})(1+O(x-c_2)),\\
&x\to c_2\,,
\end{aligned}
\\
\begin{aligned}
&v_{4\to5}(x)=\frac{1}{\sqrt[4]{\pi^2\gamma}}(\tfrac{1}{2} c_4^-\e^{-\frac{2}{3}\gamma^{3/2}}+c_4^+\e^{\frac{2}{3}\gamma^{3/2}})\\
&\times(1+O(1/\gamma^{3/2})),
\quad
\gamma\to\infty\,.
\end{aligned}
\end{gather*}
Matching these expansions, we obtain
\vspace*{-0.4ex}
\begin{eqnarray*}
   \begin{pmatrix}
     a_5^- \\
     a_5^+ \\
   \end{pmatrix}=
   C_5
   \begin{pmatrix}
    c_4^+\\
    c_4^-\\
   \end{pmatrix}.
%   C_{11}
%   = \frac{(a\epsilon)^{1/6}}{\sqrt{\pi k}}
%   \begin{pmatrix}
%   1 &\  0\\
%   0  &\  \frac{1}{2} \\
%  \end{pmatrix}.
\end{eqnarray*}

%%%%%%%%%%%%%%%%%%%%%%%%%%%%%%%%%%%%%%%%%%%%%%%%%%%%%%%%%%%%%%%%%%%%%%%%%%%%%%%%%%%%%%%%%%%%%
\paragraph{Trace of the monodromy matrix for $0<\lambda<1$.}
%
%\kern-\medskipamount
%\noindent
We now finally ready to use the results of the previous section to compute the monodromy matrix for $0<\lambda<1$.
In the sub-region $x\in(-\pi,c_1)$,
we choose the fundamental matrix solution $Y(x)$ as in Eq.~\eref{e:fundamental_matrix_solution},
with $v_{+}(x)$ and $v_{-}(x)$ given by Eq.~\eref{e:solution_region1}.
The value of this fundamental matrix solution at $x=-\pi$ is thus
\begin{equation*}
\begin{aligned}
&Y(-\pi)= \\
&\begin{pmatrix}
\begin{smallmatrix}
A_-(-\pi)\e^{-S_2(\lambda)/2\epsilon}& A_+(-\pi)\e^{S_2(\lambda)/2\epsilon}\\
\frac{1}{\epsilon}\sqrt{|Q(-\pi)|}A_-(-\pi)\e^{-S_2(\lambda)/2\epsilon}&-\frac{1}{\epsilon}\sqrt{|Q(-\pi)|}A_+(-\pi)\e^{S_2(\lambda)/2\epsilon}
\end{smallmatrix}
\end{pmatrix}\!,
\end{aligned}
\end{equation*}
with $A_\pm$ given by Eq.~\eref{e:A}.
Based on the above discussion, the value of the continuation of the above fundamental matrix solution at $x=\pi$ is given by
\begin{equation*}
\begin{aligned}
&Y(\pi)=\\
&\begin{pmatrix}
\begin{smallmatrix}A_-(\pi)\e^{S_2(\lambda)/2\epsilon}& A_+(\pi)\e^{-S_2(\lambda)/2\epsilon}\\
\frac{1}{\epsilon}\sqrt{|Q(\pi)|}A_-(\pi)\e^{S_2(\lambda)/2\epsilon}&-\frac{1}{\epsilon}\sqrt{|Q(\pi)|}A_+(\pi)\e^{-S_2(\lambda)/2\epsilon}
\end{smallmatrix}\end{pmatrix}C,
\end{aligned}
\end{equation*}
where $C$ is the connnection matrix given by
\vspace*{-1ex}
\begin{equation}
C=C_5C_8C_3C_7C_1C_6C_5C_4C_3C_2C_1\,.
\end{equation}
Now recall that, as before, we can obtain the monodromy matrix via Eq.~\eqref{e:monodromy}.
Tedious but straightforward algebra then shows that the trace of the monodromy matrix is given by Eq.~\eref{e:trM},
with $S_1(\lambda)$ and $S_2(\lambda)$ given by Eq.~\eref{e:S_1} and Eq.~\eref{e:S_2}, as before.

%%%%%%%%%%%%%%%%%%%%%%%%%%%%%%%%%%%%%%%%%%%%%%%%%%%%%%%%%%%%%%%%%%%%%%%%%%%%%%%%%%%%%%%%%%%%%
\subsection{\blue{A.3~ Calculation of relative bandwidth}}

\beginblue
% introduce a separate subsection for better reference in the text
\paragraph{Approximated expressions for $S_1(\lambda)$ and $S_2(\lambda)$.}
The functions $S_1(\lambda)$ and $S_2(\lambda)$ characterise the oscillating and the envelope part of the analytical expression of the trace of the monodromy matrix in Eq.~\eqref{e:monodromy}.
These are crucial quantities since $S_1(\lambda)$ determine the location of the bands through Eq.~\eref{e:mndef} and in turn the amplitudes and velocities of the relative soliton-like excitations,
whereas the solitonic character of the band is fixed by the condition \eref{e:lambdasdef} which depends on $S_2(\lambda)$.
In order to obtain explicit estimates of the number of solitons along with their features
(amplitudes and velocities) as a function of $\epsilon$, it is worth to introduce appropriate Taylor expansions.
\blue{Specifically,} the Taylor expansions of $S_1(\lambda)$ around $\lambda=0$ and of $S_2(\lambda)$ around $\lambda=1$
\blue{are needed.}
\blue{Using known properties of elliptic integrals \cite{NIST}, one can verify that these expansions}
are given by the following expressions
\vspace*{-0.2ex}
\bse
\begin{gather}
S_1(\lambda) = \pi\lambda/2 + \sqrt{2}\pi\lambda^2/16 + O(\lambda^3)\,,
\label{e:s1Taylor}
\\
S_2(\lambda)=-\pi(\lambda-1)/2+O(\lambda-1)^2\,,\quad
\end{gather}
\ese
respectively.
\endblue

%The Taylor expansions of $S_1(\lambda)$ around $\lambda=0$ and of $S_2(\lambda)$ around $\lambda=1$ are given by
%\vspace*{-1ex}
%\begin{gather}
%\begin{aligned}
%&S_1(\lambda) = \pi\lambda/2 + \sqrt{2}\pi\lambda^2/16 + O(\lambda^3)\,,\\
%&S_2(\lambda)=-\pi(\lambda-1)/2+O(\lambda-1)^2\,,
%\end{aligned}
%\end{gather}
%respectively. These expressions are used to obtain explicit estimates of the number of solitons as well as their amplitudes and velocities as a function of $\epsilon$.

\paragraph{Relative bandwidth.}
%
%\kern-\medskipamount
%\noindent
Here we give some details on the characterization of the bands.
For $n\geq1$, the width of the $n$-th spectral band, which is approximately centered at $z_n$,
is given by $w_n=\lambda_{2n}-\lambda_{2n-1}$,
while for $n=0$, the bandwidth is given by $w_0=\lambda_0$. The width of the $n$-th spectral gap is given by $g_n=\lambda_{2n+1}-\lambda_{2n}$.
%\textcolor{blue}{Recall the more relevant quantity, i.e.,
%relative bandwidth $W_n$ is defined in Eq.~\eref{e:wnWn} in the main text.}
%To characterize $W_n$,
We first consider the Taylor expansion of $\tr M$ around $z_n$ (the location of the $n$-th maximum),
evaluated at $\lambda=\lambda_{2n-1}$ and $\lambda=\lambda_{2n}$,
%Because of the exponentially large oscillations of the trace function, it is sufficient to only keep quadratic terms.
%By doing so we can approximate the condition $\tr M=-2$ as
%$-2 = (\tr M)''(z_n)(z_n-\lambda_{2n-2})^2/2+O(z_n-\lambda_{2n-2})^3= (\tr M)''(z_n)(z_n-\lambda_{2n-1})^2/2+O(z_n-\lambda_{2n-1})^3$.
%Combining these expressions and recalling~\eref{e:trM},
and we obtain the $n$-th bandwidth as
\vspace*{-0.2ex}
\begin{equation*}
w_n=\frac{2\epsilon}{|S_1'(z_n)|}\exp(-S_2(z_n)/\epsilon)
+O(\epsilon \e^{-2S_2(z_n)/\epsilon})\,.
\end{equation*}

Next we use the difference $z_{n+1}-z_n$ to approximate $w_n+g_n$.
To do so, we first show the difference between these two quantities is given by
\vspace*{-0.2ex}
\be
\nonumber
(w_n+g_n)-(z_{n+1}-z_n)=O(\epsilon \e^{-S_2(z_{n+1})/\epsilon}).
\ee
%\begin{equation}
%\begin{aligned}
%&(w_n+g_n)-(z_{n+1}-z_n)\\
%= &\frac{1}{2}(w_n - w_{n+1})+O(\epsilon \e^{-2S_2(z_{n+1})/\epsilon})\\
%=& O(\epsilon \e^{-S_2(z_{n+1})/\epsilon}).
%\end{aligned}
%\end{equation}
%Also note that
%$S_1(z_{n+1}) - S_1(z_n) = \pi\epsilon$.
%and $S_1(\lambda_n)=\frac{2n-1}{4}\pi\epsilon$.
We then expand $S_1(\lambda)$ around $z_n$, evaluate at $\lambda=z_{n+1}$ and obtain
\vspace*{-1ex}
\be
\nonumber
z_{n+1}-z_n = \pi\epsilon/S_1'(z_n)+O(\epsilon^2).
\ee
%Combining these results, we then obtain
%\be
%w_n+g_n=\frac{\pi\epsilon}{S_1'(z_n)}+O(\epsilon^2).
%\label{e:wngn}
%\ee
Combining all the above results, %and \textcolor{blue}{using the equivalent expression $W_n = w_n/(w_n + g_n)$,}
we finally obtain the asymptotic expression for the $n$-th relative bandwidth in Eq.~\eref{e:Wn}.

%%%%%%%%%%%%%%%%%%%%%%%%%%%%%%%%%%%%%%%%%%%%%%%%%%%%%%%%%%%%%%%%%%%%%%%%%%%%%%%%%%%%%%%%%%%%%
%\input references

\end{document}